\documentclass{aa}
\usepackage{graphicx}
\usepackage{epsfig,graphicx,rotating,portland,fancyheadings,longtable}
\usepackage{multirow}
\usepackage{txfonts}

\def\gs{{_>\atop^{\sim}}}

\begin{document}
\title{VLT/X-shooter spectroscopy of the GRB\,090926A afterglow
  \thanks{Based on observations collected at the European Southern
    Observatory, ESO, the VLT/Kueyen telescope, Paranal, Chile, during
 the science verification phase, proposal code: 060-9427(A).}}

\author{V. D'Elia$^{1,2}$, J. P. U. Fynbo$^{3}$, S. Covino$^{4}$,
  P. Goldoni$^{5,6}$, P. Jakobsson$^{7}$, F. Matteucci$^{8}$,
  S. Piranomonte$^{1}$, J. Sollerman$^{3,9}$, C.C. Th\"one$^{4}$,
  S.D. Vergani$^{10,11}$, P.M. Vreeswijk$^{3}$, D.J. Watson$^{3}$,
  K. Wiersema$^{12}$, T. Zafar$^{3}$, A. de Ugarte Postigo$^{4}$,
  H. Flores$^{11}$, J. Hjorth$^{3}$, L. Kaper$^{13}$,
  A.J. Levan$^{14}$, D. Malesani$^{3}$, B. Milvang-Jensen$^{3}$,
  E. Pian$^{8,15}$, G. Tagliaferri$^{4}$, N.R. Tanvir$^{12}$
}

\institute
{$^1$INAF-Osservatorio Astronomico di Roma, Via Frascati 33, I-00040 Monteporzio Catone, Italy\\
$^2$ASI-Science Data Centre, Via Galileo Galilei, I-00044 Frascati, Italy\\
$^3$Dark Cosmology Centre, Niels Bohr Institute, University of Copenhagen, Juliane Maries Vej 30, 2100 Copenhagen, Denmark  \\
$^{4}$INAF-Osservatorio Astronomico di Brera, via E. Bianchi 46, 23807 Merate (LC), Italy\\  
$^5$Laboratoire Astroparticule et Cosmologie, 10 rue A. Domon et L. Duquet, 75205 Paris Cedex 13, France\\
$^6$DSM/IRFU/Service D'Astrophysique, CEA-Saclay, 91191 Gif-sur-Yvette, France\\
$^7$Centre for Astrophysics and Cosmology, Science Institute, University of Iceland, Dunhagi 5, 107 Reykjavik, Iceland \\
$^8$INAF, Osservatorio Astronomico di Trieste, Via Tiepolo 11, 34143 Trieste, Italy \\
$^9$The Oskar Klein Centre, Department of Astronomy, AlbaNova, Stockholm University, 106 91 Stockholm, Sweden \\
$^{10}$University Paris 7, APC, Lab. Astroparticule et Cosmologie, UMR7164 CNRS, 10 rue Alice Domon et LÂonie Duquet 75205 Paris Cedex 13, France  \\
$^{11}$Laboratoire Galaxies Etoiles Physique et Instrumentation, Observatoire de Paris, 5 place Jules Janssen, 92195 Meudon, France \\
$^{12}$Department of Physics and Astronomy, University of Leicester, Leicester LE1 7RH, UK \\
$^{13}$Astronomical Institute ``Anton Pannekoek'', University of Amsterdam, Science Park 904, 1098 XH Amsterdam, The Netherlands \\
$^{14}$Department of Physics, University of Warwick, Coventry, CV4 7AL, UK \\ 
$^{15}$Scuola Normale Superiore, Piazza dei Cavalieri 7, 56126, Pisa, Italy;\\
}

  \abstract 
  {} 
  {The aim of this paper is to study the environment
    and intervening absorbers of the gamma-ray burst GRB 090926A through analysis of optical
spectra of its afterglow.}
{We analyze medium resolution spectroscopic observations ($R=10 000$,
  corresponding to 30 km s$^{-1}$, S/N$=15 - 30$ and wavelength range
  $3 000-25 000$) of the optical afterglow of GRB\,090926A, taken with
  X-shooter at the VLT $\sim 22$ hr after the GRB trigger.}
  {The spectrum shows that the ISM in the GRB host galaxy at $z =
    2.1071$ is rich in absorption features, with two components
    contributing to the line profiles. In addition to the ground state
    lines, we detect {\ion{C}{II}}, {\ion{O}{I}}, {\ion{Si}{II}},
    {\ion{Fe}{II}} and {\ion{Ni}{II}} excited absorption
    features. No host galaxy emission lines, molecular absorption features
    nor diffuse interstellar bands are detected in the
    spectrum. The line of sight of GRB\,090926A presents four weak intervening
    absorption systems in the redshift range $ 1.24 < z < 1.95$.}
{The Hydrogen column density associated to GRB\,090926A is $\log
  N_{\rm H}/{\rm cm}^{-2} = 21.60 \pm 0.07$, and the metallicity of the
  host galaxy is in the range [X/H] $=
  3.2\times10^{-3}-1.2\times10^{-2}$ with respect to the solar values,
  i.e., among the lowest values ever observed for a GRB host galaxy. A
  comparison with galactic chemical evolution models has suggested
  that the host of GRB090926A is likely to be a dwarf irregular
  galaxy. No emission lines were detected, but we note that a
  H$\alpha$ flux in emission of $9\times10^{-18}$
  erg~s$^{-1}$~cm$^{-2}$ (i.e., a star formation rate of
  $2~M_\odot$yr$^{-1}$), which is typical of many GRB hosts, would
  have been detected in our spectra, and thus emission lines are well
  within the reach of X-shooter.
  We put an upper limit to the \ion{H}{} molecular fraction of the host
  galaxy ISM, which is $f < 7\times10^{-7}$.  The continuum has been
  fitted assuming a power-law spectrum, with a spectral index of
  $\beta = 0.89^{+0.02}_{-0.02}$. The best fit does not essentially
  require any intrinsic extinction since $E_{B-V} < 0.01$ mag
  adopting a SMC extinction curve.

  We derive information on the distance between the host absorbing gas
  and the site of the GRB explosion.  The distance of component I is
  found to be $2.40 \pm 0.15$ kpc, while component II is located far
  away from the GRB, possibly at $\sim 5$ kpc. These values are
  compatible with that found for other GRBs.

}

\authorrunning {D'Elia et al.}
\titlerunning {X-shooter spectroscopy of GRB\,090926A afterglow}

\maketitle
%

\section{Introduction}

The study of the inter stellar medium (ISM) of $z\gs1$ galaxies has
traditionally relied upon Lyman-break galaxies (LBGs) at $z=3-4$ (see
e.g. Steidel et al.  1999), K-band selected galaxies (Savaglio et
al. 2004) and galaxies which happen to be along the lines of sight to
bright background quasars (or QSOs).  However, LBGs are characterized
by pronounced star-formation and their inferred chemical abundances
may relate to these regions rather than being representative of
typical high-redshift galaxies.  Weak metal line systems along the
line of sight to quasars probe mainly galaxy haloes, rather than their
bulges or discs (Fynbo et al. 2008). Taking advantage of ultra-deep
Gemini multi-object spectrograph observations, Savaglio et al. (2004,
2005) studied the ISM of a sample of faint $K$-band selected galaxies
at $1.4 < z < 2.0$, finding MgII and FeII abundances much higher than
in QSO systems but similar to those in GRB host galaxies. Such studies
can hardly be extended to higher redshift with the present generation
of 8m class telescopes, because of the faintness of high-redshift
galaxies. Since the discovery that gamma-ray bursts (GRBs) are
extragalactic, we now can avail of an independent tool to study the
ISM of high-redshift galaxies.


Metallicities measured in GRB host galaxies vary from less than
$10^{-2}$ to nearly solar values and are on average larger than
those found along QSO sightlines (see e.g., Fynbo et al. 2006,
Prochaska et al. 2007). This result supports the notion that GRBs
originate in dwarf and/or low-mass galaxies.
Since star formation occurs in molecular clouds, the latter are
expected to be the GRB birthplaces. In this scenario, absorption from
ground-state and vibrationally excited levels of the H$_2$ molecules
are expected, but these are typically not observed in GRB afterglow
spectra (Vreeswijk et al. 2004, Tumlinson et al. 2007). The
non-detection of H$_2$ molecules (with the exception of GRB\,080607,
see Prochaska et al. 2009, Sheffer et al. 2009) can possibly be
explained by photo-dissociation by the intense UV flux from the
GRB afterglow.

Indeed, the main difference between QSO and GRB absorption
spectroscopy is that QSOs are nearly stationary in their emission,
while GRBs are the most variable and violent phenomena in the
Universe.  Thus, while QSOs have the time to ionize the ISM along
their lines of sight, the physical, dynamical and chemical status of
the circumburst medium in the star-forming region hosting GRB
progenitors can be modified by the explosive event, through shock
waves and ionizing photons. The transient nature of GRBs is manifested
through the detection of fine structure and other excited levels of
the atom {\ion{O}{I}} and the ions {\ion{Fe}{II}}, {\ion{Ni}{II}},
{\ion{Si}{II}} and {\ion{C}{II}}.  These features are routinely
identified in GRB spectra, and are most probably excited by the
intense UV flux from the afterglow, since strong variation is observed
when multi-epoch spectroscopy is available.  This variation is not
consistent with a pure infrared excitation or collisional processes
(Prochaska, Chen \& Bloom 2006, Vreeswijk et al. 2007, D'Elia et
al. 2009a). Thus, assuming UV pumping as the responsible mechanism for
the production of these lines, the distance of the gas to the GRB can
be computed. This distance comes out to be of the order of a few
hundred pc (see D'Elia et al. 2009b for GRB\,080330 and Ledoux et
al. 2009 for GRB\,050730) or even in the kpc scale (see Vreeswijk et
al. 2007 for GRB\,060418 and D'Elia et al. 2009a for the naked-eye
GRB\,080319B).

GRB spectroscopy is also suitable for studying systems lying along the
line of sight to GRBs. 
Surprisingly, the number density of strong {\ion{Mg}{II}} intervening
absorbers in GRB spectra is more than twice larger than along QSO
sightlines (Prochter et al. 2006, Vergani et al. 2009), while
{\ion{C}{IV}} absorbers do not show any statistical difference
(Sudilovsky et al. 2007, Tejos et al. 2007). The reason for the
{\ion{Mg}{II}} excess in GRB spectra is still unclear, and a larger
sample is needed to properly address this issue (Porciani et al. 2007,
Cucchiara et al. 2009).

All these issues can now be systematically addressed using the
X-shooter spectrograph.  This is the first second-generation
instrument at the ESO's Very Large Telescope (VLT) at Paranal
Observatory (Chile). It is a single target spectrograph capable of
obtaining a medium resolution spectrum ($R=\lambda/\Delta\lambda=4000
- 14,000$) covering the spectral range 3000 - 24,800~\AA{} in a single
exposure thanks to the splitting of the light into three arms:
ultraviolet/blue (UVB), visual (VIS) and near-infrared (NIR). The high
spectrograph efficiency allowed us to obtain good quality observations
of the GRB\,090313 afterglow (de Ugarte Postigo et al. 2010), although
it being observed two days after the burst and under unfavourable
conditions. We refer the reader to D'Odorico et al. (2006) for a more
complete description of the X-shooter specifications and capabilities.

We discuss here the case of GRB\,090926A, observed by X-shooter the
27th of September 2009.  We investigate both the local medium
surrounding the GRB, and the intervening systems. We derive the
metallicity by comparing the column densities of Hydrogen and metals,
we search for molecular absorption and galactic emission lines, and we
constrain the distance between the GRB and the absorber by comparing
the ratios between the ground and excited level column densities with
photoexcitation codes, under the assumption of indirect UV pumping
production of the excited levels. An analysis of the intervening
absorbers lying along the line of sight to the GRB afterglow and a
search for diffuse interstellar bands (DIBs) is also presented.

The paper is organized as follows. Section $2$ makes a short summary
of the GRB\,090926A detection and observations from the literature;
Section $3$ presents the X-shooter observations and data reduction;
Sections $4$ and $5$ are devoted to the study of the features from the
host galaxy, Sect. $6$ derives the extinction curve shape for this
GRB; Section $7$ presents the analysis of the other absorbing systems
identified in the GRB\,090926A line of sight; finally in Sect. $8$ the
results are discussed and conclusions are drawn. We assume a cosmology
with $H_0=70$ km s$^{-1}$ Mpc$^{-1}$, $\Omega_{\rm m} = 0.3$,
$\Omega_\Lambda = 0.7$. Hereafter, the [] refers to element abundances
relative to Solar values and X refers to any chemical element.

\section{GRB\,090926A}

GRB\,090926A was discovered by {\it Fermi} on September 26, 2009, at
04:20:26 UT, and was detected by both the GBM (Bissaldi 2009) and the
LAT (Uehara et al. 2009) instruments. {\it Swift} repointed to the target
13 hr later and found the X-ray counterpart (Vetere et
al. 2009). The afterglow was later reported to be detected at optical
wavelengths with the Skynet/PROMPT telescopes (Haislip et al. 2009)
and with UVOT (Gronwall \& Vetere 2009); the reported PROMPT
magnitude was $R\sim18$, although the observations took place almost 20
hr post burst. The redshift was secured by X-shooter, which
observed the afterglow 2 hr after these optical detections. The
preliminary reported value was $z=2.1062$ (Malesani et al. 2009). FORS2
spectroscopic observations and GROND photometric follow-up of the
GRB\,090926A afterglow have been presented in Rau et al. (2010,
hereafter R10): we will compare our results with those of R10 in the
following.


\begin{figure}
\centering
\includegraphics[angle=-0,width=9cm]{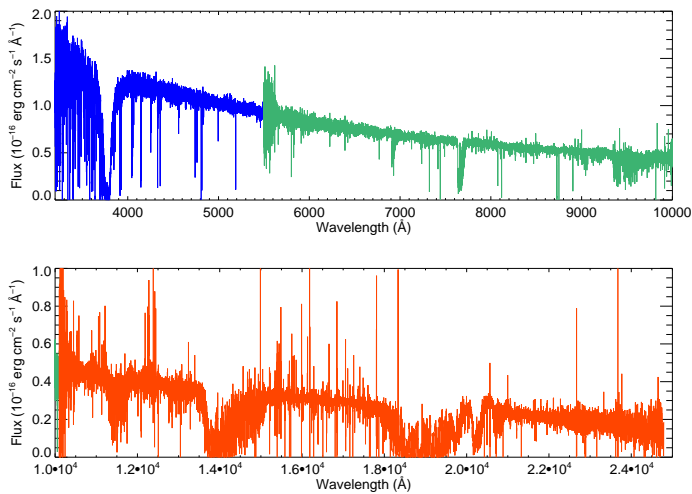}
\caption{The flux-calibrated X-shooter spectrum of
  GRB\,090926A. Top: ultraviolet/blue and visual arms. Bottom:
  near-infrared arm. The absolute flux calibration may be not
accurate due to slit losses.}
\label{spe1}
\end{figure}


\begin{table}[ht]
\caption{\bf X-shooter observations}
\centering
{\footnotesize
\smallskip
\begin{tabular}{|l|c|c|c|}
\hline 
Observation & Time since burst (hr) & Exposure (s) &  S/N       \\
\hline                          
1           & 22.13                 & 600          &  $5-15$    \\
\hline                          
2           & 22.29                 & 600          &  $5-15$    \\
\hline                            
3           & 22.51                 & 600          &  $5-15$    \\
\hline                          
4           & 22.68                 & 600          &  $5-15$    \\
\hline                         
1+2+3+4     & 22.37                 & 2400         &  $10-30$   \\
\hline

\end{tabular}
}
\end{table}


\section{Observations and data reduction}

In the framework of the Science Verification phase program, we
observed the afterglow of GRB\,090926A with X-shooter (D'Odorico et
al. 2006), mounted at the VLT-UT2 telescope. The observations consist
of 4 different exposures of 600 s each (see Table 1). The exposures
were taken using the nodding along the slit technique with an offset
of 5 arcsec between exposures in a standard ABBA sequence. The
sequence began on September 27th, 2007 at 02:23:14 UT, $\sim 22$~hr
after the GRB trigger. The magnitude of the afterglow at the time of
the observation was reported by R10 to be $R = 18.7$, which translates
into a $6600$\AA \, flux of $0.8 \times 10^{-16}$ erg cm$^{-2}$
s$^{-1}$ \AA $^{-1}$. The total net exposure time of our observations
is 40 min. The slit width was set to 0.9\arcsec{} in the VIS and NIR
arms and 1.0\arcsec{} in the UVB arm. The UVB and VIS CCD detectors
were rebinned to $1 \times 2$ pixels (binned in the spectral direction
but not in the spatial one) to reduce the readout noise.

We processed the spectra using a preliminary version of the X-shooter
data reduction pipeline (Goldoni et al. 2006). The pipeline performed
the following actions: The raw frames were first bias subtracted and
cosmic ray hits were detected and removed using the method developed
by van Dokkum (2001). The orders were extracted and rectified in
wavelength space using a wavelength solution previously obtained from
calibration frames.  The resulting rectified orders were shifted and
coadded to obtain the final 2D spectrum. In the overlapping regions,
orders were merged by weighting them according to the errors
propagated during the entire reduction process. From the resulting 2D
merged spectrum, a one dimensional spectrum was extracted at the
source position. The one dimensional spectrum with the corresponding
error file and bad pixel map is the final product of the reduction.

The resolution is $R \sim 10,000$ and the achieved spectral range is
$\sim 3000$ to $\sim 24,800$~\AA. The data below $\sim 3020$ and above
$\sim 24,000$~\AA{} are, however, totally dominated by noise. To
perform flux calibration we extracted a spectrum from a staring
observation of the flux standard BD +17${\deg}$4708 (Bohlin \&
Gilliland 2004).  In this case we subtracted the sky emission lines
using the Kelson (2003) method.  This spectrum was divided by the flux
table of the same star from the CALSPEC HST database (Bohlin 2007,
http://www.stsci.edu/hst/observatory/cdbs/calspec.html) to produce the
response function.  The response was then interpolated where needed in
the atmospheric absorption bands in VIS and NIR and applied to the
spectrum of the source. No telluric correction was applied, so that
prominent atmospheric bands, especially in the NIR arm, can still be
seen (Fig.~1). We searched for variability in the absorption features,
but we found none (the equivalent widths in the four spectra are
constant at the $1.5 \sigma$ level), so we summed the four,
flux-calibrated observations (see Fig.~1). This lack of variability is
not surprising, since the spectra have been acquired nearly 1 day
after the burst (see Sect. 4.3 for details). The signal-to noise ratio
(per pixel) ranges from $\sim 10$ to $\sim 30$ in the co-added
spectrum.  In order to perform line fitting (see section 4) the
spectrum was normalized to the continuum, which was evaluated by
fitting the data with cubic splines, after the removal of the
absorption features. Finally, the noise spectrum, used to determine
the errors on the best fit line parameters, was calculated from the
real, background-subtracted spectrum using line-free regions. This
takes into account both statistical and systematic errors in the
pipeline processing and background subtraction.


\section{The host galaxy atomic absorption features}

The gas residing in the GRB host galaxy is responsible for many of the
features observed in the GRB\,090926A afterglow spectrum.  Metallic
features are apparent, from neutral ({\ion{O}{I}}, {\ion{Mg}{I}},
{\ion{Ca}{I}}), low ionization ({\ion{C}{II}}, {\ion{Mg}{II}},
{\ion{Al}{II}}, {\ion{Al}{III}}, {\ion{Si}{II}}, {\ion{S}{II}},
{\ion{Ca}{II}}, {\ion{Fe}{II}}, {\ion{Ni}{II}}) and high ionization
({\ion{C}{IV}}, {\ion{N}{V}}, {\ion{O}{VI}}, {\ion{Si}{IV}},
{\ion{S}{IV}}) species. In addition, strong absorption from the fine
structure levels of {\ion{C}{II}}, {\ion{O}{I}}, {\ion{Si}{II}},
{\ion{Fe}{II}} and from the metastable level of {\ion{Ni}{II}} is
identified, suggesting that the intense radiation field from the GRB
excites such features. Table~2 gives a summary of all the absorption
lines due to the host galaxy gas. The analysis of the spectral
features has been performed with FITLYMAN (Fontana \& Ballester
1995). This program is able to simultaneously fit several absorption
lines, linking the redshifts, column densities and Doppler parameters
if required. FITLYMAN takes into account the atomic masses when a
thermal model of the Doppler broadening is adopted (as we did),
enabling to link the Doppler parameters of different species. The
probed ISM of the host galaxy is resolved into two components
separated by $48$ km s$^{-1}$ which contribute to the absorption
system. The wealth of metal-line transitions allows us to precisely
determine the redshift of the GRB host galaxy. This yields a
vacuum-heliocentric value of $z=2.1071 \pm 0.0001$, setting the
reference point to the red component of the main system; this
component is the only one for which there is significant absorption
from all the excited levels of the intervening gas (see Sect.~4.1 and
4.3). This redshift value supersedes that reported by Malesani et
al. (2009), which was based on archival calibration data and using an
older X-shooter pipeline version. The host galaxy environment will be
described in the next sub-sections together with a study of the
excited lines aimed at estimating the distance from the GRB of the
circumburst gas.

\subsection{Line fitting procedure}

In general, the analysis of the GRB environment is not straightforward due to
the complexity of the absorption lines profile, which in
several cases cannot be fitted with a single line profile.
This means that many components contribute to the gas in the GRB
environment.  In other words, several layers of gas which may be close
to or far from each other, appear mixed together in the spectrum (in
velocity space).  The presence of several components is thus
indicative of clumpy gas in the GRB environment, composed of different
absorbing regions each with different physical properties. This
behaviour is particularly evident for GRB afterglows observed at high
resolution (see e.g. Fiore et al. 2005, Th\"one et al. 2008).

For what concerns GRB\,090926A, the {\ion{C}{IV}} and {\ion{Si}{IV}}
lines have the wider velocity range. This behaviour is common to many
GRBs (see e.g. D'Elia et al. 2007, Piranomonte et al. 2008), where
these lines are the ones that are most clearly detected and used to
guide the identification of the different components constituting the
circumburst matter.  A two-component model provides a good fit for the
{\ion{C}{IV}} and {\ion{Si}{IV}} lines (Fig.~2). Thus, the redshifts
and the Doppler parameters of these components were fixed in order to
fit the other species present in the spectrum with the same
model. This modeling adequately fits all other absorption lines at the
GRB redshift.  All species feature absorption in the red component,
and most of them have absorption also in the blue component.  Figures
3 and 4 show the two-component model fit to all the absorption lines
at the GRB redshift. Asterisks mark the fine structure levels. The
lines {\ion{Ni}{II}~$\lambda\lambda$2166, 2217 and 2316} (Fig.~4,
left) are produced by the \ion{Ni}{II} second excited level
($^4F_{9/2}$).  The {\ion{Si}{III}} blue component could not be
derived because of blending with Ly$\alpha$ forest features (Fig.~4, right).

The column densities for all the elements and ions of the host galaxy
absorbing gas, estimated using this two-component model, are reported
in Table 2. The upper limits reported are at the 90\% confidence
level. The red and blue components are marked as I and II,
respectively.  The reference zero point of the velocity shifts has
been placed at $z = 2.1071$, coincident with the redshift of the red
component. Component I of {\ion{Al}{III}} needs a slight wavelength
red-shift in order to be adequately fitted (Fig. 3, bottom
left). Anyway, the resulting column density of this component is
consistent with the value obtained linking the {\ion{Al}{III}} and
{\ion{C}{IV}} central wavelengths. The features of some species, such
as \ion{Mg}{II}, \ion{O}{I}, \ion{O}{VI} and \ion{C}{II}, could be
saturated at least in component I. In this case, the column densities
in Table 2 should be regarded as lower limits to the real values. The
column density of \ion{Si}{II} has been calculated using the
$\lambda$1304 transition only (see Sect. 4.3). It is interesting to
note that among the excited absorption features in the spectrum,
produced by fine structure and/or metastable levels, \ion{C}{II},
\ion{Si}{II} and \ion{Ni}{II} show significant absorption in both
components I and II (the absorption in component II from \ion{Ni}{II}
is marginal), while \ion{O}{I} and \ion{Fe}{II} only show this in the
red, reference component I. We fixed the zero point to the redshift of
component I because the \ion{Fe}{II} fine structure levels require a
high UV flux in order to remain populated (see Sect.~4.3), implying
that component I is possibly closer to the GRB explosion site than
II. As described in the introduction, the observation of excited
states in GRB absorption spectra is a quite common
feature. Sub-section 4.3 is thus devoted to these features and to the
information that can be extracted from their analysis.
 
\begin{table}[ht]
\caption{\bf Absorption line column densities for the two components of the main system.}
{\footnotesize
\smallskip
\begin{tabular}{|lc|cc|}
\hline
Species              & Observed transitions                        & I (0 km s$^{-1}$) & II ($-48$ km s$^{-1}$)  \\
\hline
\ion{H}{I}  $^2S_{1/2}$        & Ly$\alpha$, Ly$\beta$                      & $21.60 \pm 0.07$  & -              \\
\hline
\ion{C}{II} $^2P^{0}_{1/2}$    &  $\lambda$1036, $\lambda$1334             & $14.48 \pm 0.04$  &$14.04 \pm 0.04$\\
\hline
\ion{C}{II} $^2P^{0}_{3/2}$    &  $\lambda$1037, $\lambda$1335             & $>14.45 \pm 0.04$  &$14.01 \pm 0.05$\\
\hline
\ion{C}{IV} $^2S_{1/2}$       &  $\lambda$1548, $\lambda$1550              & $14.43 \pm 0.02$  &$13.89 \pm 0.03$\\
\hline
\ion{N}{V} $^2S_{1/2}$        &  $\lambda$1238, $\lambda$1242              & $14.08 \pm 0.03$  &$13.66 \pm 0.07$\\
\hline
\ion{O}{I} $^3P_{2}$          &  $\lambda$1039, $\lambda$1302              & $>14.75 \pm 0.03$  & $ < 13.8      $\\
\hline
\ion{O}{I} $^3P_{1}$          &  $\lambda$1304                             & $14.49 \pm 0.03$  & $ < 13.7      $\\
\hline
\ion{O}{I} $^3P_{0}$          &  $\lambda$1306                             & $14.37 \pm 0.04$  & $ < 13.7      $\\
\hline
\ion{O}{VI} $^2S_{1/2}$       &  $\lambda$1031, $\lambda$1037              & $>14.60 \pm 0.16$  & Blend          \\
\hline
\ion{Mg}{I} $^1S_0$          &  $\lambda$2852                              & $12.74 \pm 0.01$  &$11.98 \pm 0.05$\\
\hline
\ion{Mg}{II} $^2S_{1/2}$      &  $\lambda$1239, $\lambda$1240               & $>14.05 \pm 0.01$  &$13.05 \pm 0.02$\\
                     &  $\lambda$2796, $\lambda$2803               &                   &                \\   
\hline
\ion{Al}{II} $^1S_0$         &  $\lambda$1670                               & $13.20 \pm 0.03$  &$12.66 \pm 0.04$\\
\hline
\ion{Al}{III} $^2S_{1/2}$     &  $\lambda$1854, $\lambda$1862               & $13.24 \pm 0.03$  &$ < 12.3      $\\
\hline
\ion{Si}{II} $^2P^{0}_{1/2}$  &  $\lambda$1020, $\lambda$1190                & $14.41 \pm 0.03$  &$13.98 \pm 0.07$\\
                     &  $\lambda$1193, $\lambda$1260                &                   &                \\   
                     &  $\lambda$1304, $\lambda$1526                &                   &                \\   
                     &  $\lambda$1808                               &                   &                \\   
\hline
\ion{Si}{II} $^2P^{0}_{3/2}$  &  $\lambda$1194, $\lambda$1197                & $13.96 \pm 0.03$  &$13.42 \pm 0.09$\\
                     &  $\lambda$1264, $\lambda$1309                &                   &                \\  
                     &  $\lambda$1533, $\lambda$1309                &                   &                \\  
\hline
\ion{Si}{III} $^1S_{0}$       &  $\lambda$1206                              & $13.50 \pm 0.11$  & Blend \\
\hline
\ion{Si}{IV} $^2S_{1/2}$      &  $\lambda$1393, $\lambda$1402               & $13.97 \pm 0.03$  &$13.61 \pm 0.04$\\
\hline
\ion{S}{II} $^4S^{0}_{3/2}$   &  $\lambda$1250, $\lambda$1253               & $14.89 \pm 0.06$  & $ < 14.3      $\\
                     &  $\lambda$1259                              &                   &                \\  
\ion{S}{IV} $^4S^{0}_{3/2}$   &  $\lambda$1062, $\lambda$1253               & $14.45 \pm 0.19$  & $ < 14.2      $\\
\hline
\ion{Ca}{I} $^1S_{0}$        &  $\lambda$4227                               & $12.15 \pm 0.07$  & $ < 11.9     $ \\
\hline
\ion{Ca}{II} $^2S_{1/2}$     &  $\lambda$3969                               & $13.21 \pm 0.03$  & $ < 12.8     $ \\
\hline
\ion{Mn}{II} $a^7S_{3}$      &  $\lambda$2576                               & $ < 12.3      $  & $ < 12.3      $\\
\hline
\ion{Fe}{II} $a^6D_{9/2}$    &  $\lambda$1081,  $\lambda$1096               & $14.03 \pm 0.03$  &$13.46 \pm 0.04$\\
                    &  $\lambda$1144,  $\lambda$2344               &                   &                \\
                    &  $\lambda$2348,  $\lambda$2374               &                   &                \\
                    &  $\lambda$2382,  $\lambda$2586               &                   &                \\
                    &  $\lambda$2600                               &                   &                \\
\hline
\ion{Fe}{II} $a^6D_{7/2}$    &  $\lambda$2333,  $\lambda$2396               & $12.52 \pm 0.08$  & $ < 12.2      $\\
                    &  $\lambda$2612,  $\lambda$2626               &                   &                \\
\hline
\ion{Fe}{II} $a^4F_{9/2}$    &  $\lambda$1559,  $\lambda$2332               & $ < 13.6       $  & $ < 13.6      $\\
\hline
\ion{Fe}{III} $^5D_{4}$    &  $\lambda$1122                                 & $14.54 \pm 0.07$  &$14.38 \pm 0.10$\\
\hline
\ion{Ni}{II} $^2D_{5/2}$     &  $\lambda$1317,  $\lambda$1370               & $13.51 \pm 0.06$  &$13.09 \pm 0.30$\\
                    &  $\lambda$1454,  $\lambda$1703               &                   &                \\
                    &  $\lambda$1709,  $\lambda$1741               &                   &                \\
                    &  $\lambda$1751                               &                   &                \\
\hline
\ion{Ni}{II} $^4F_{9/2}$     &  $\lambda$2166,  $\lambda$2217               & $13.54 \pm 0.02$  &$12.47 \pm 0.36$\\
                    &  $\lambda$2316                               &                   &                \\
\hline
\end{tabular}

All values of the column densities are logarithmic (in cm$^{-2}$). Reported lower limits are due to possible line saturation.
}
\end{table}

\begin{figure}
\centering
\includegraphics[angle=-0,width=9cm]{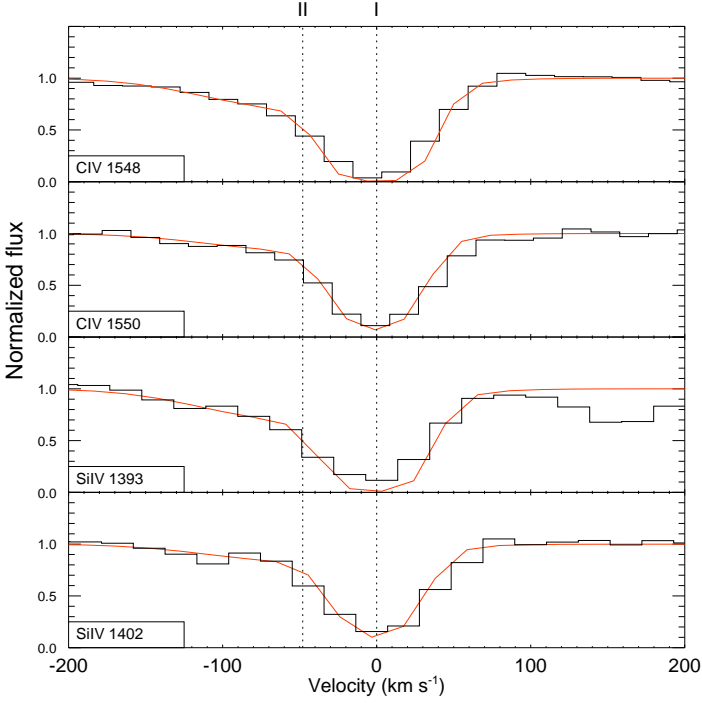}
\caption{The \ion{C}{IV} and \ion{Si}{IV} absorption features. Solid
  lines represent the two Voigt components, best fit model. Vertical
  lines identify the component velocities. The
  zero point has been arbitrarily placed at the redshift of the red
  component ($z=2.1071$).  }
\label{spe1}
\end{figure}

\begin{figure}
\centering
\includegraphics[angle=-0,width=4.4cm,height=7.5cm]{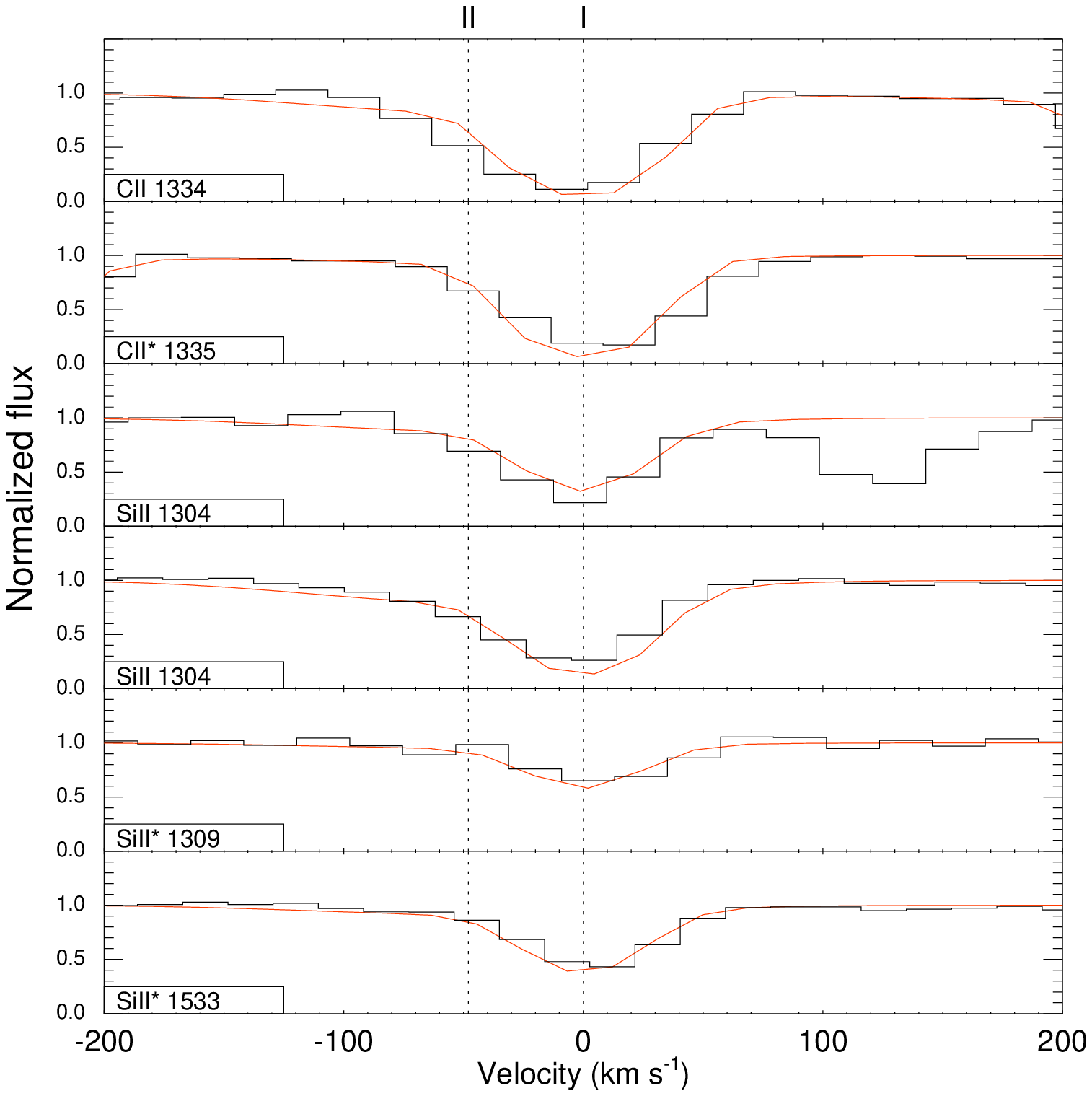}
\includegraphics[angle=-0,width=4.4cm,height=7.5cm]{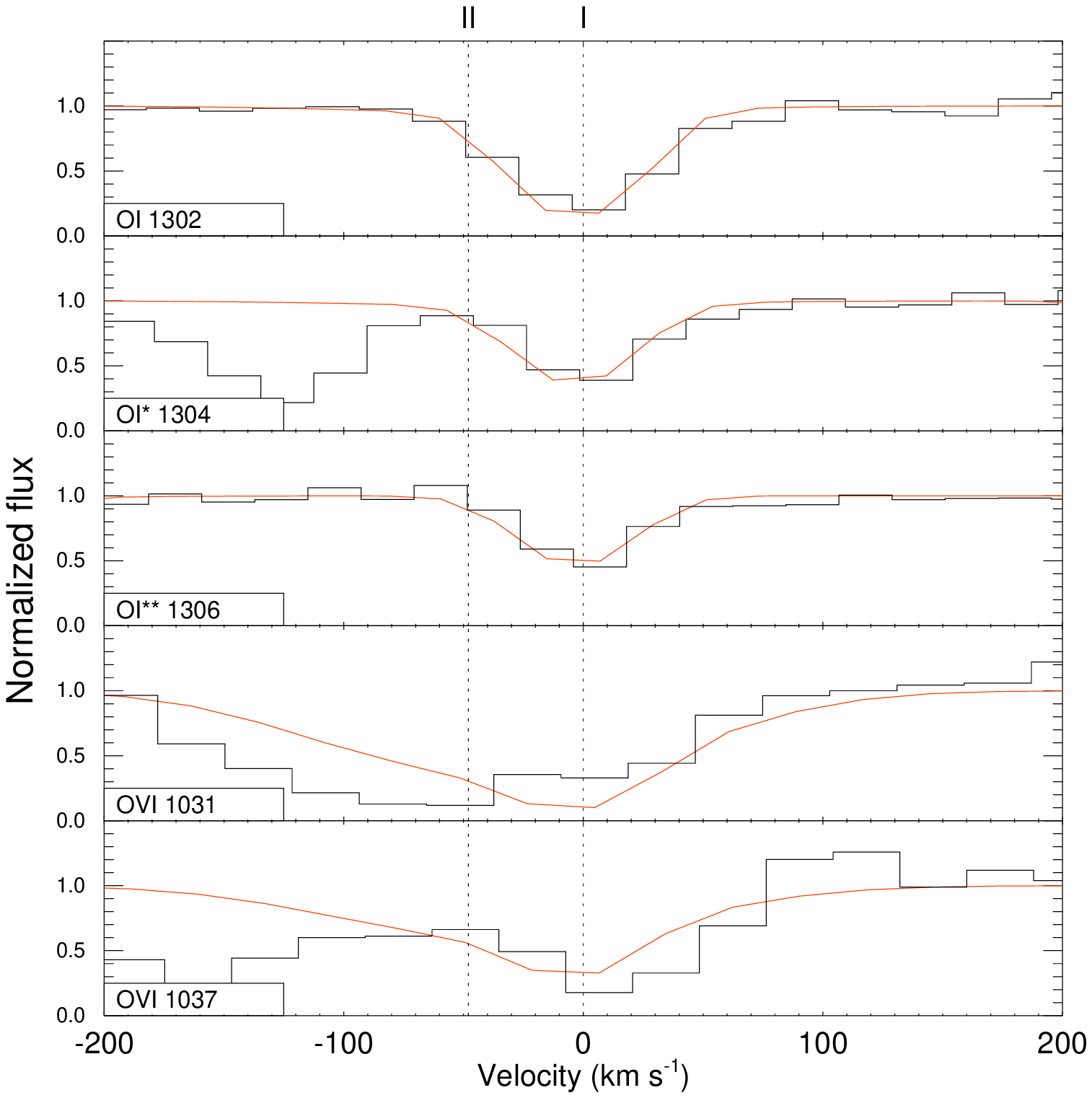}
\includegraphics[angle=-0,width=4.4cm,height=7.5cm]{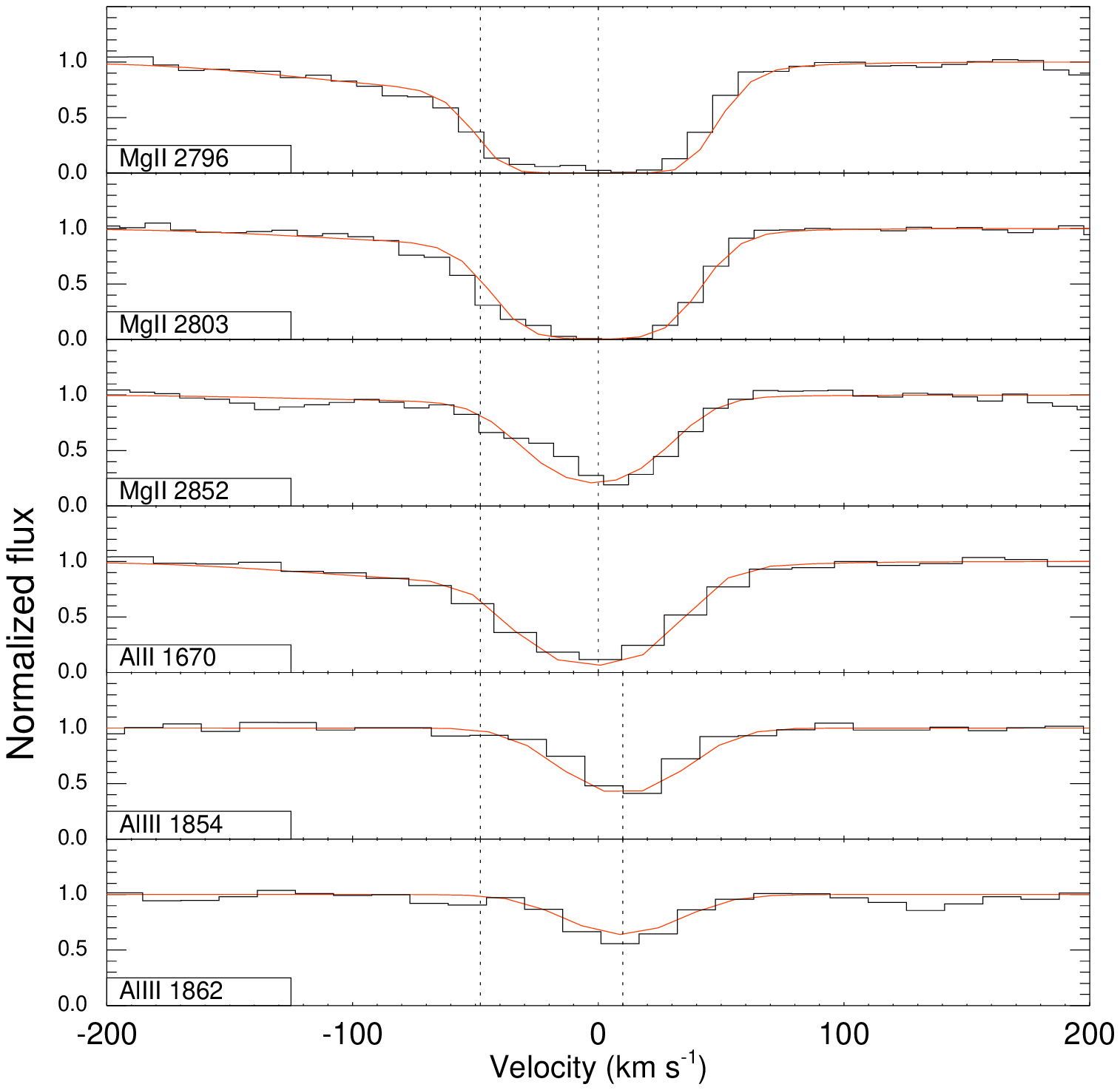}
\includegraphics[angle=-0,width=4.4cm,height=7.5cm]{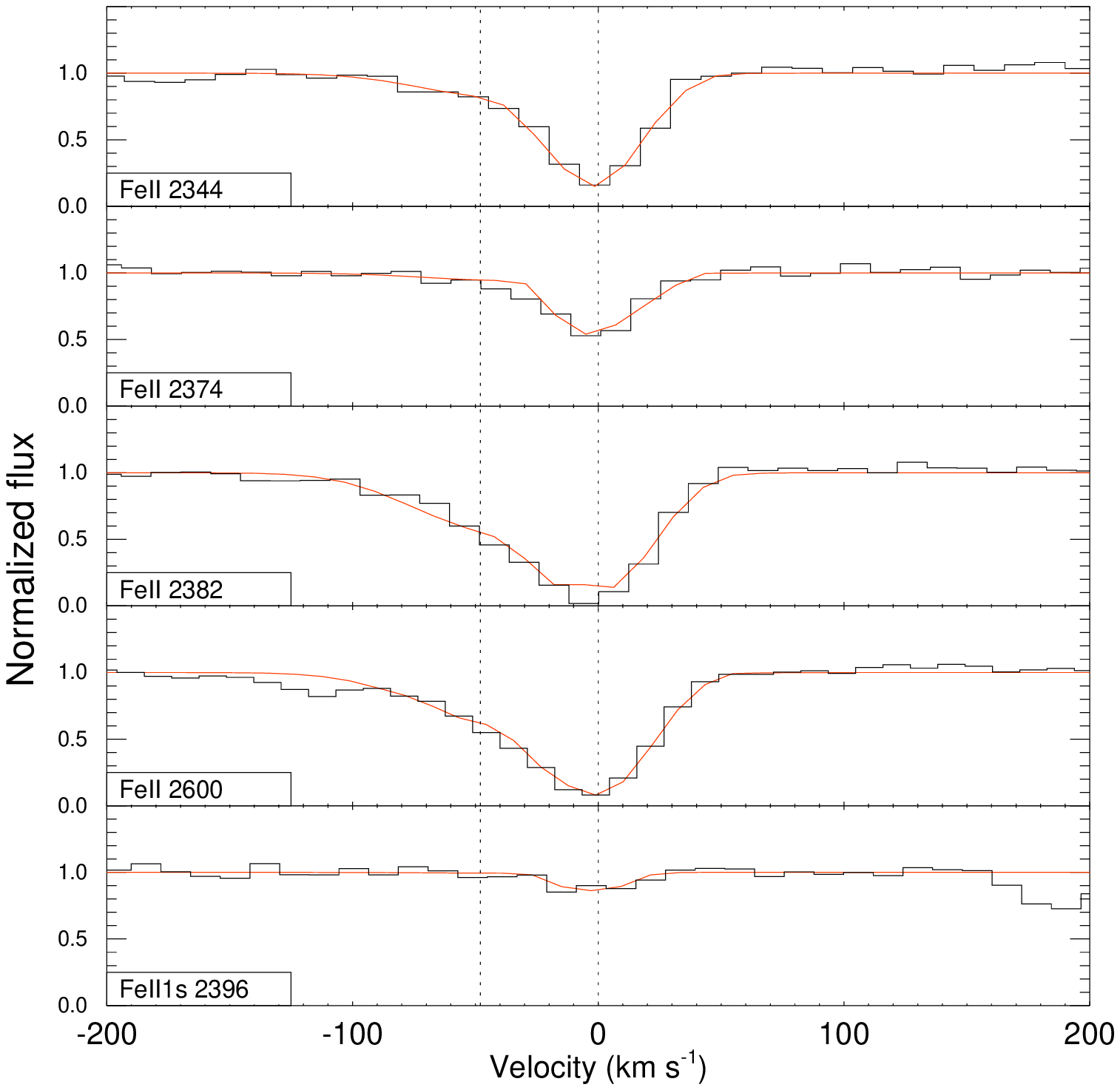}
\caption{The \ion{C}{II} and \ion{Si}{II} (top left panel), \ion{O}{I}
  and \ion{O}{VI} (top right panel), \ion{Mg}{I}, \ion{Mg}{II},
  \ion{Al}{II} and \ion{Al}{III} (bottom left panel), \ion{Fe}{II}
  (bottom right panel) absorption features. Solid lines represent the
  two Voigt components, best fit model. Vertical lines identify the
  component velocities. The zero point has been arbitrarily placed at
  the redshift of the red component ($z=2.1071$). The \ion{O}{VI}
  features are blended at blue wavelengths. Component I of
  \ion{Al}{III} has been slightly shifted in order to be adequately
  fitted.}
\label{spe1}
\end{figure}

\begin{figure}
\centering
\includegraphics[angle=-0,width=4.4cm,height=7.5cm]{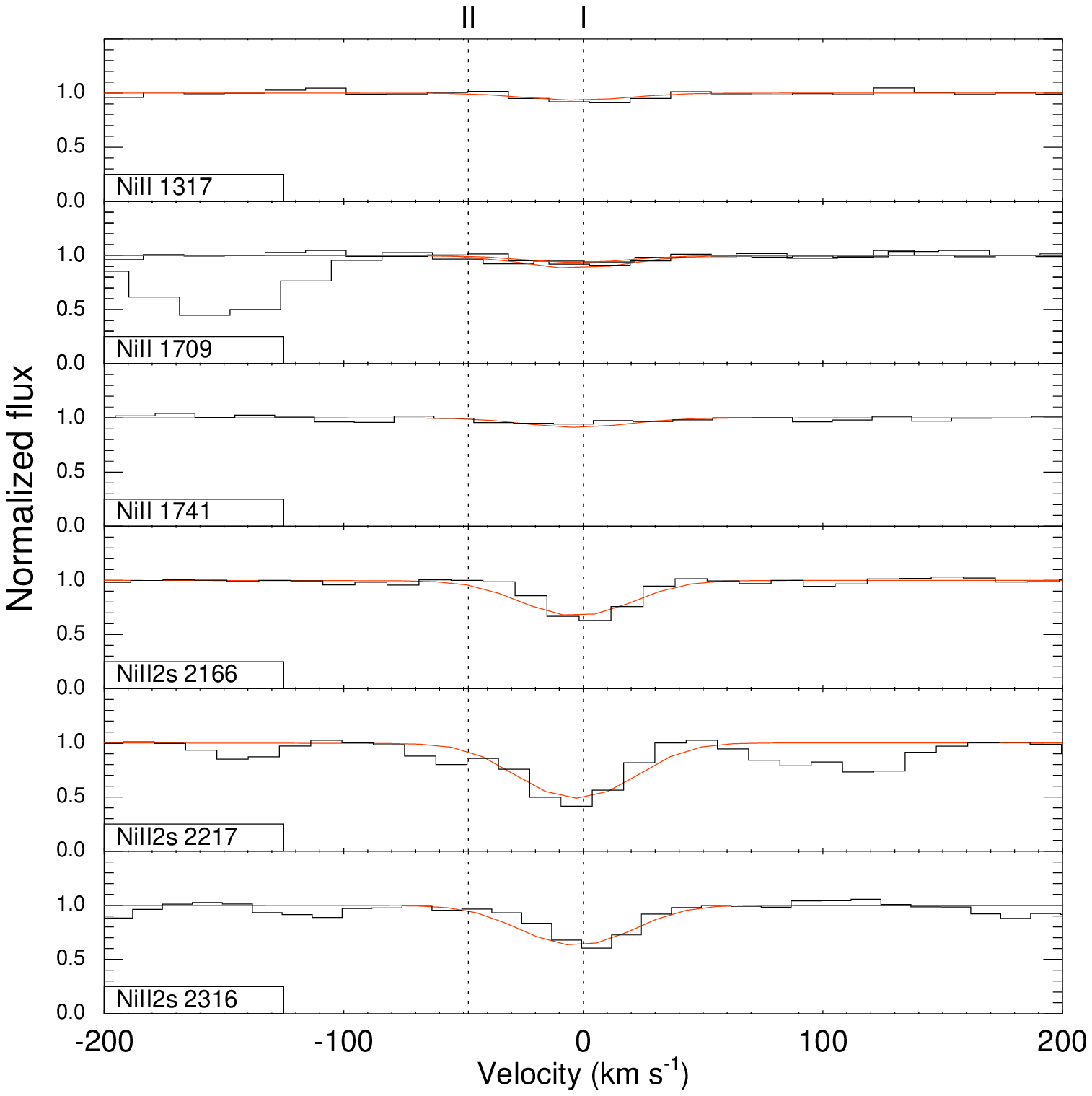}
\includegraphics[angle=-0,width=4.4cm,height=7.5cm]{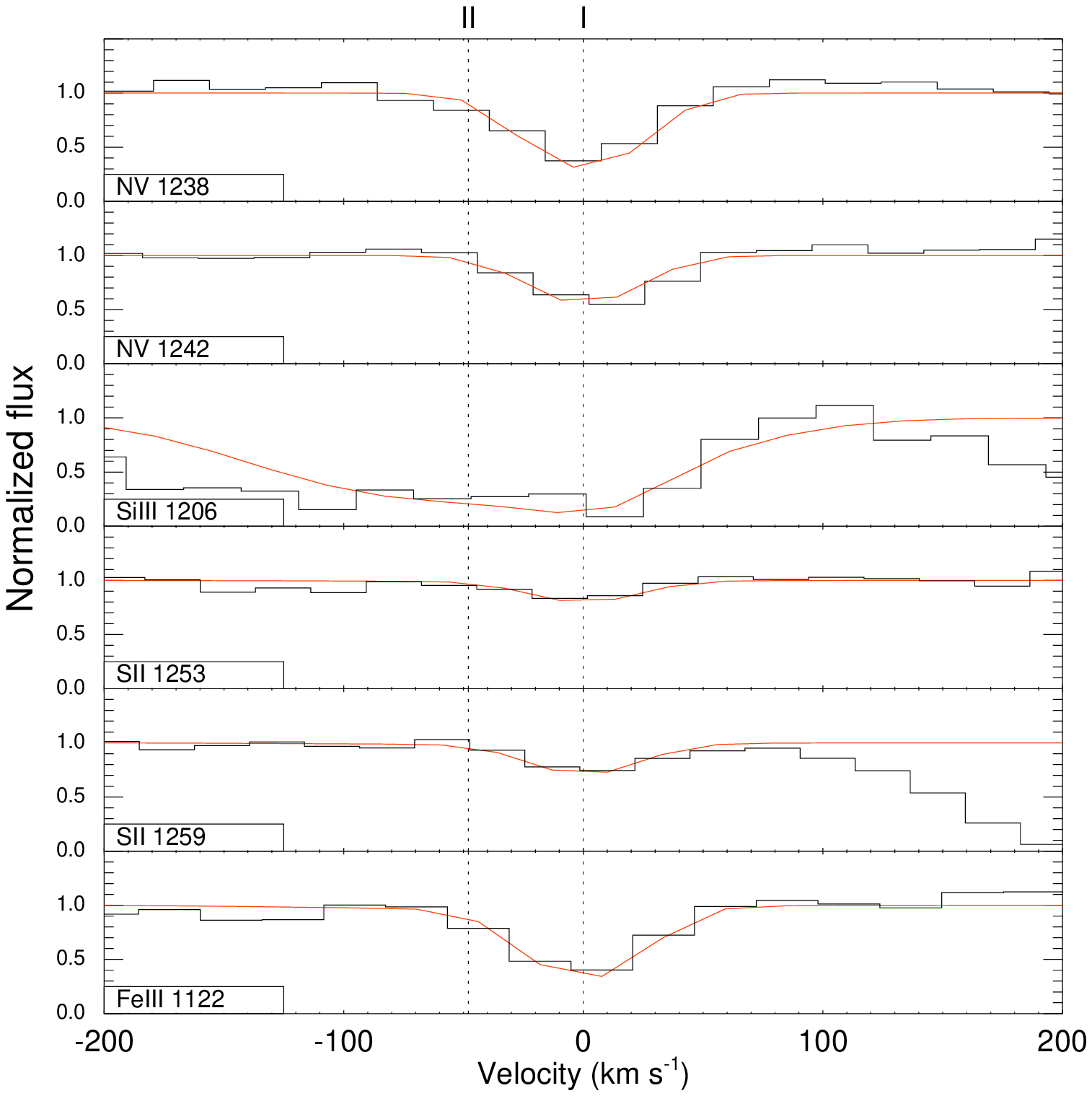}
\caption{Left panel: the \ion{Ni}{II} absorption features. Right
  panel: the \ion{N}{V}, \ion{Si}{III}, \ion{S}{II} and \ion{Fe}{III}
  absorption features. Solid lines represent the two Voigt components,
  best fit model. Vertical lines identify the component
  velocities. The zero point has been arbitrarily placed at the
  redshift of the red component ($z=2.1071$). The blue component of
  \ion{Si}{III} is blended with absorption from the Ly$\alpha$
  forest.}
\label{spe1}
\end{figure}

\subsection{Metallicities}

The GRB\,090926A redshift was high enough to allow the Hydrogen
Ly$\alpha$ and Ly$\beta$ lines to enter the X-shooter spectral
window. Ly$\alpha$ can be clearly seen in Fig.~1 (upper panel) at
$\sim 4000$ \AA. We used the two Lyman features in order to constrain
the Hydrogen column density. Figure~5 shows the Lyman-$\alpha$ and
Lyman-$\beta$ profiles. Both features are damped and we can not
separate the two components identified in the metal-line fits. The
Hydrogen column density computed has $\log (N_{\rm H}/{\rm cm}^{-2}) =
21.60 \pm 0.07$ cm$^{-2}$, for a reduced $\chi^2$ value of $1.33$, and
is virtually insensitive to the adopted b parameter. From this $N_{\rm
  H}$ value, we can compute the metallicity for the host of
GRB\,090926A, using the metallic column densities reported in Table
2. Due to the wide spectral coverage of X-shooter, this can be done
for a large number of elements. We proceeded as follows: first of all,
since the Lyman features cannot be resolved into components, we summed
up line by line the values in table 2 (e.g., the two components) to
obtain the total column density for each species; second, we summed
the total column densities of the transitions belonging to the same
atom (different ionization and excitation states), in order to
evaluate the atomic column densities. These values have been divided
by $N_{\rm H}$, and compared to the corresponding solar values given
in Asplund et al. (2009). The results are listed in Table 3. Column 2
reports the total abundance of each atom, while columns 3 and 4 report
the absolute and solar-scaled $N_X$/$N_{\rm H}$ ratios, respectively,
with $X$ being the corresponding element in column 1. Lower limits are
reported whenever saturation does not allow us to securely fit the
metallic column densities. We derive very low metallicity values with
respect to the solar ones, between $4.2\times10^{-3}$ and
$1.4\times10^{-2}$. The very low value derived for \ion{N} is due to
the fact that we could not fit low ionization lines, but the
\ion{N}{V} species only, meaning that this value is not truly
representative for all the N ionization states.

\begin{figure}
\centering
\includegraphics[angle=-0,width=8.3cm,height=6.5cm]{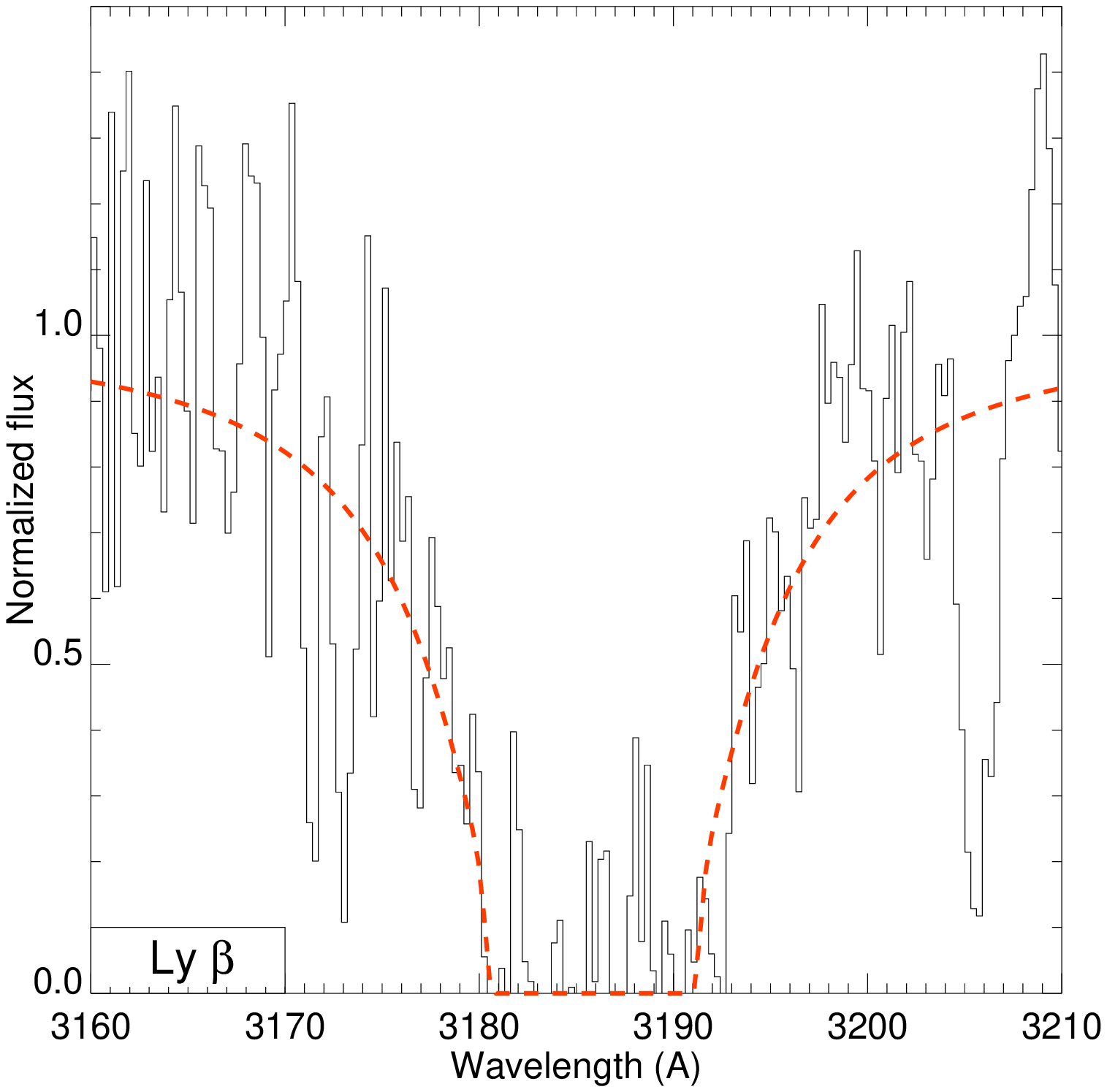}
\includegraphics[angle=-0,width=9.cm]{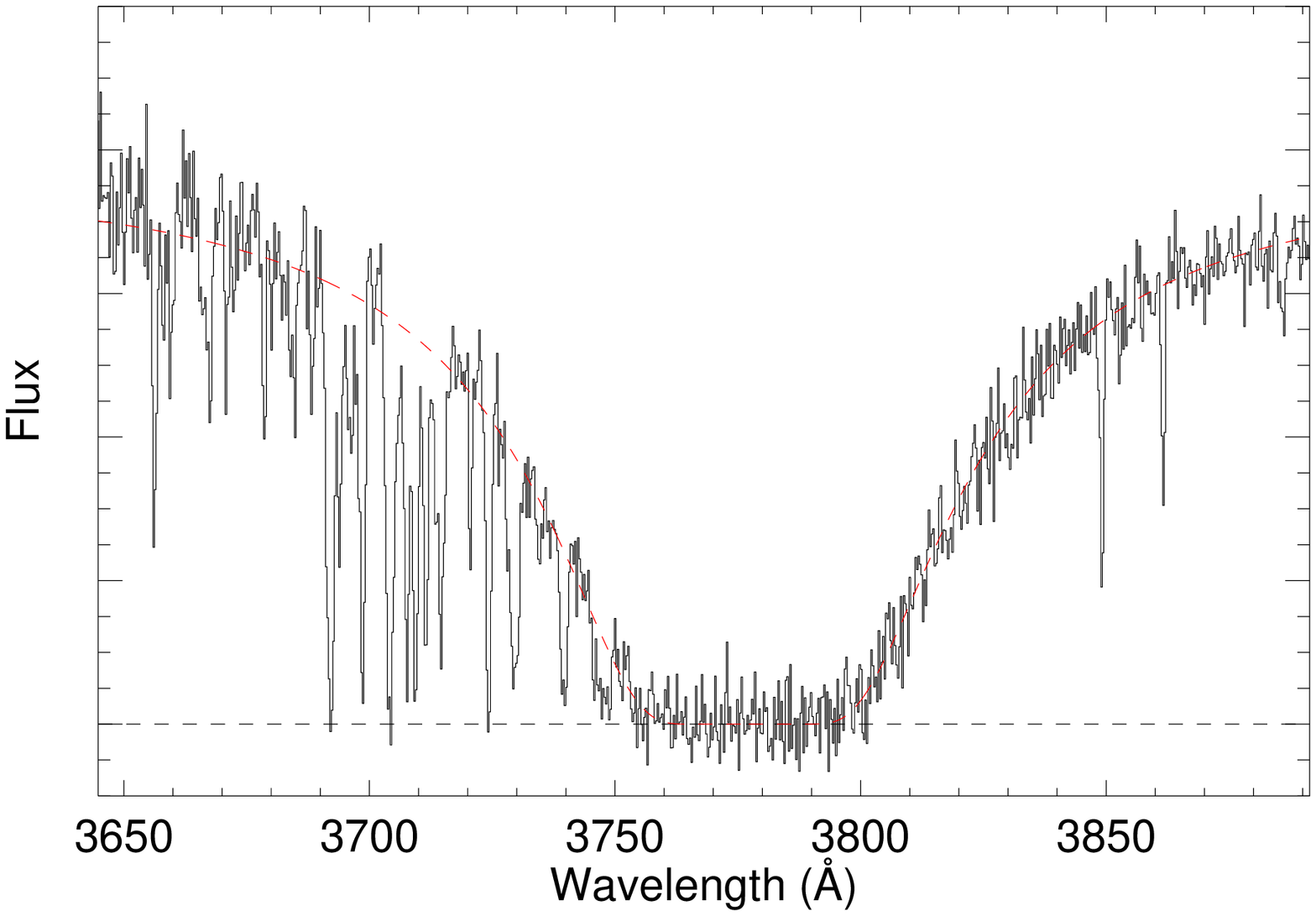}
\caption{The Ly$\beta$ (top panel) and Ly$\alpha$ (bottom panel)
  absorption features at the GRB\,090926A redshift. The dashed lines
  represent the single Voigt component, best fit model, centered at
  the redshift of the red component ($z=2.1071$) of the
  metallic lines.}
\label{spe1}
\end{figure}


\begin{table}[ht]
\caption{\bf Metallicities}
{\footnotesize
\smallskip
\begin{tabular}{|l|ccc|}
\hline
Element $X$& $\log N_X /{\rm cm}^{-2}$   &$\log N_X$/$N_{\rm H}$    & $[X/{\rm H}]$    \\
\hline
C      & $>15.05\pm0.10$  & $>-6.55\pm0.10$ & $>-2.94\pm0.10$ \\
N      & $ 14.22\pm0.08$  & $ -7.38\pm0.08$ & $ -3.16\pm0.08$ \\
O      & $>15.31\pm0.10$  & $>-6.29\pm0.10$ & $>-2.95\pm0.10$ \\
Mg     & $>14.11\pm0.08$  & $>-7.49\pm0.08$ & $>-3.02\pm0.08$ \\
Al     & $ 13.59\pm0.10$  & $ -8.01\pm0.10$ & $ -2.38\pm0.10$ \\
Si     & $ 14.80\pm0.08$  & $ -6.80\pm0.08$ & $ -2.31\pm0.08$ \\
S      & $ 14.89\pm0.10$  & $ -6.71\pm0.10$ & $ -1.85\pm0.10$ \\
Ca     & $ 13.25\pm0.08$  & $ -8.35\pm0.08$ & $ -2.66\pm0.08$ \\                      
Fe     & $ 14.86\pm0.09$  & $ -6.74\pm0.09$ & $ -2.19\pm0.09$ \\
Ni     & $ 13.92\pm0.13$  & $ -7.68\pm0.13$ & $ -1.91\pm0.13$ \\
\hline
\end{tabular}
}
\end{table}

\subsection{Excited levels}

The level structure of an atom or ion is characterized by a principal
quantum number $n$, which defines the atomic level, and by the
spin-orbit coupling (described by the quantum number $j$), which
splits these levels into fine structure sub-levels. In GRB absorption
spectra, several excited features are detected at the GRB redshift,
due to the population of both $n>1$ and/or $n=1$ fine structure
levels. As mentioned before, component I of the main system in the
spectrum of GRB\,090926A shows several absorptions from excited states. In
particular, the first fine structure level of the \ion{Fe}{II}
ground state ($a^6D$), the \ion{Si}{II} $^2P^{0}_{3/2}$,
{\ion{C}{II}} $^2P^{0}_{3/2}$, \ion{O}{I} $^3P_{1}$ and \ion{O}{I}
$^3P_{0}$ fine structure levels and the {\ion{Ni}{II}}
$^4F_{9/2}$ metastable level are present. Moreover, excited states of
\ion{Si}{II}, \ion{C}{II} and (marginally) \ion{Ni}{II} are detected
also in component II (see Table 2 for details).

There are basically two mechanisms to excite the gas of the GRB host
galaxy to such states.  The first is through collisional effects (if
the electron density is sufficiently high, i.e., $\ge 10^5$
cm$^{-3}$); the second is through the absorption of electromagnetic
radiation. In the latter case, the absorbed photons can be infrared,
through direct population of the fine-structure levels of the ground
state and other excited levels with a low value of $n$, or UV, through
the population of higher levels followed by the de-population into the
states responsible for the absorption features. Multi-epoch
spectroscopy together with proper modeling of the atomic level
population has proven to be a powerful tool to discriminate between
these two processes. The strong variability in the column density of
the {\ion{Fe}{II}} and {\ion{Ni}{II}} excited levels observed in
GRB\,060418 (Vreeswijk et al. 2007) and GRB\,080319B (D'Elia et
al. 2009a) ruled out collisional processes and direct infrared pumping
as being responsible for the excitation. A collisional origin of the
excitation can be ruled out even if multi-epoch spectroscopic data are
missing. In fact, collisions populate higher energy levels less than
lower energy ones. For instance, GRB\,080330 (D'Elia et al. 2009b) and
GRB\,050730 (Ledoux et al. 2009) exhibit a \ion{Fe}{II} $a^4F_{9/2}$
excited state column density larger than that of several fine
structure levels of the \ion{Fe}{II} $a^6D$ ground state. This means
that radiative processes are at work also for these GRBs. For
GRB\,090926A we do have multi-epoch spectroscopy, but the time lag
between different observations (a few minutes, see Table 1) is too
short when compared to the time delay between the GRB explosion and
the epoch at which our data were taken. Supported by these results on
atomic excitation mechanisms, we assume that indirect UV pumping by
the fading afterglow is at work also for GRB\,090926A.

As a first step, we assume a steady state flux coming from the GRB,
which in general is not a good approximation. Nevertheless, the
GRB\,090926A afterglow was observed by X-shooter almost one day after
the GRB explosion, and its flux level was decaying slowly at this
stage. Under this assumption, we compute the ratio between the first
excited level and the ground state for \ion{Si}{II}, \ion{O}{I} and
\ion{Fe}{II}. We then use these values as input for the plot in Fig. 7
of Prochaska, Chen \& Bloom (2006) in order to estimate the flux
experienced by the absorbing gas. This plot has been produced using
the model from Silva \& Viegas (2002), which further assumes optically
thin clouds. We first note that for component I the \ion{Fe}{II} and
\ion{Si}{II} ratios are not compatible, i.e., the flux levels inferred
from these ions are not consistent. This is possily due to a slight
saturation of the \ion{Si}{II} $\lambda$1526 line (see Fig. 3, top
left), which was formerly used together with the $\lambda$1304
transition to evaluate the \ion{Si}{II} column density.  Therefore, we
recomputed the \ion{Si}{II} column density using just the \ion{Si}{II}
$\lambda$1304 line, as described in Sect. 4.1. The flux levels with
the new $N_{\ion{Si}{II}}$ for component I are now fully
compatible. We further note that the \ion{O}{I} ratio is far from
being compatible with that of \ion{Si}{II} and \ion{Fe}{II}. This
confirms the saturation of the \ion{O}{I}~$\lambda$1302 transition
(Fig. 3, middle), which can just be used to set a lower limit to the
\ion{O}{I} ground state column density. The \ion{Si}{II} ratio for
component II is close to that for I. Thus, we derive a distance of the
gas responsible for the absorption of the two components I and II of
$d = 1.8 \pm 0.2$ kpc.


\begin{figure}
\centering
\includegraphics[angle=-0,width=8.5cm]{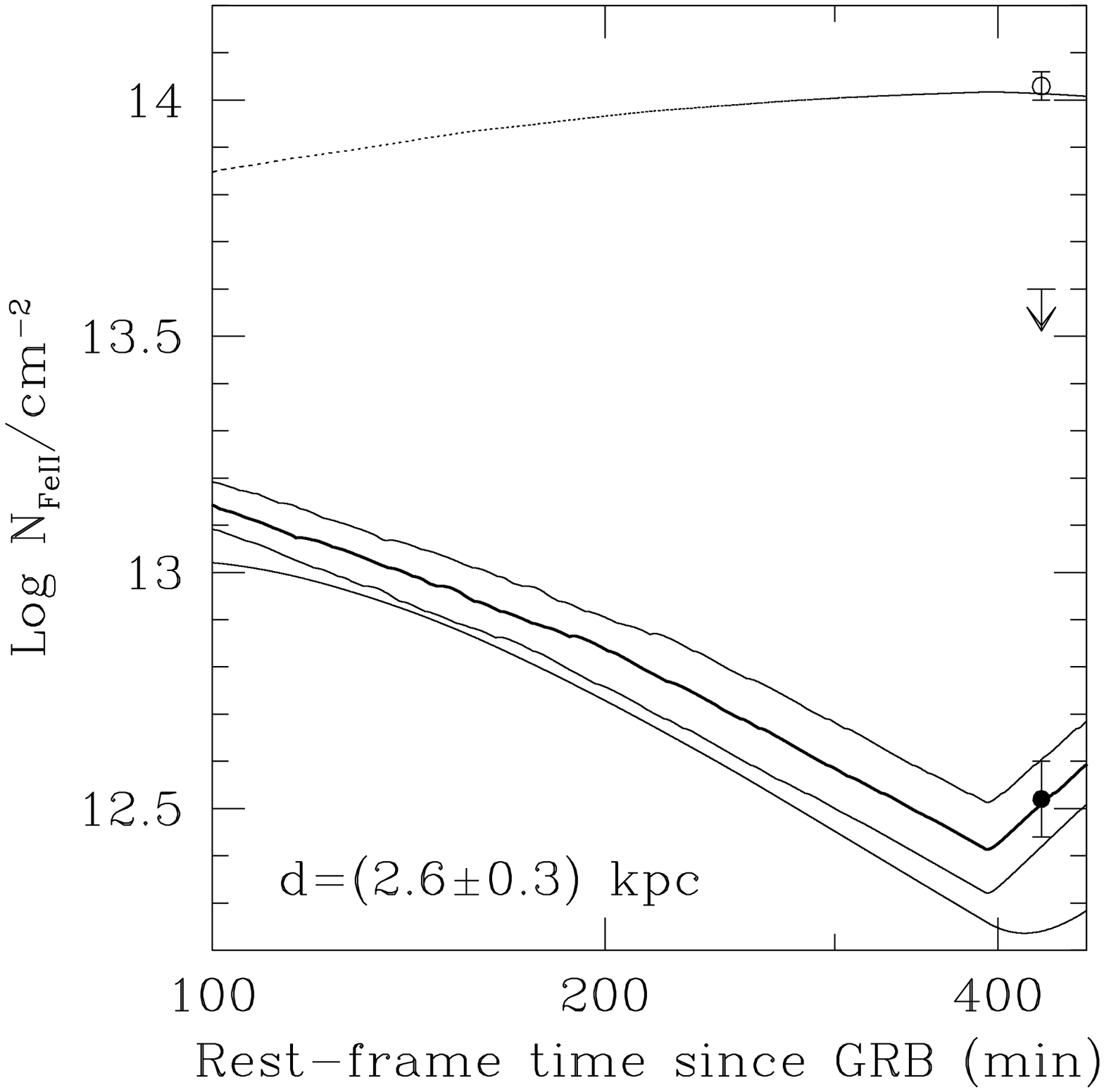}
\includegraphics[angle=-0,width=8.5cm]{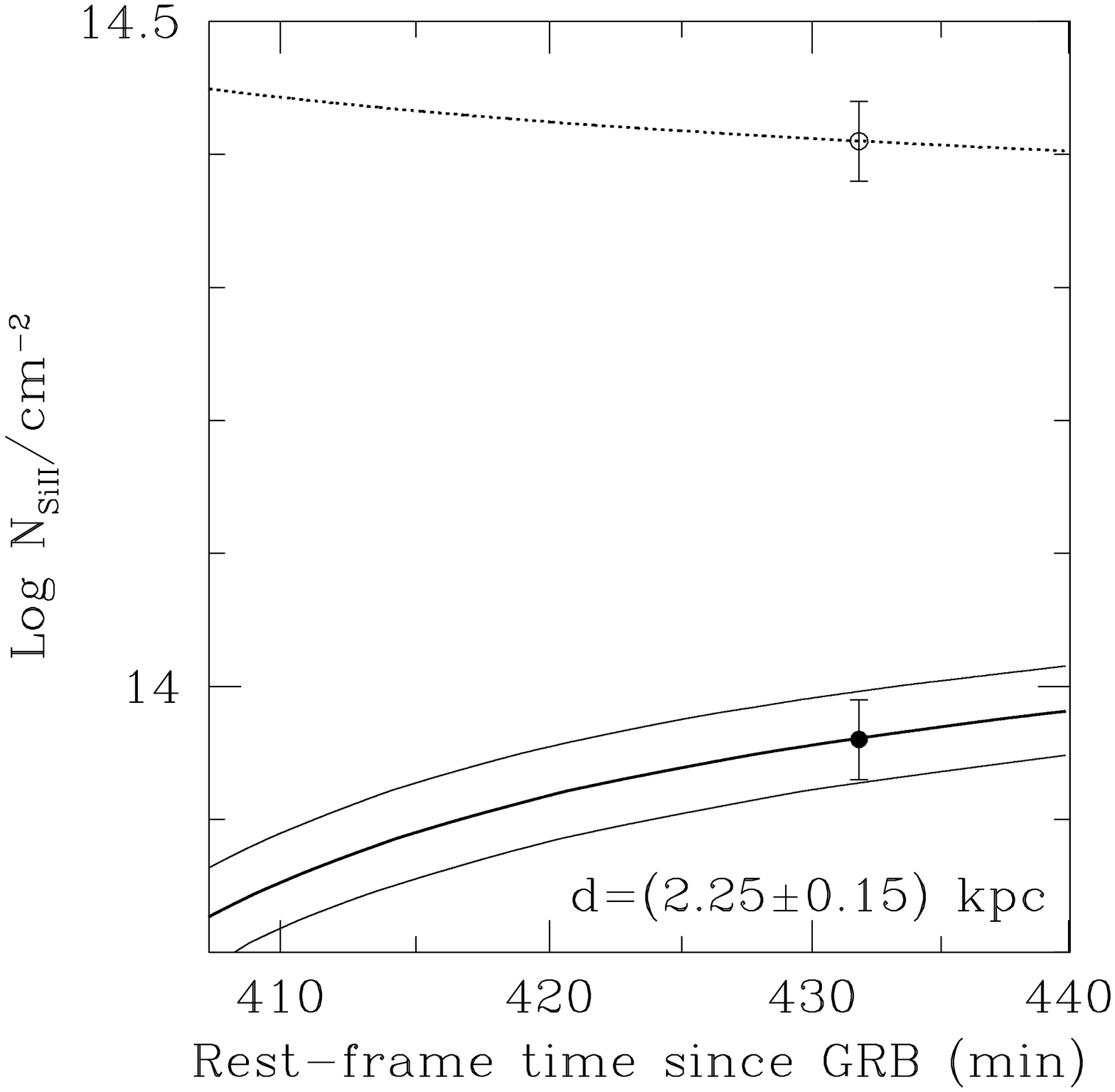}
\caption{Top panel: \ion{Fe}{II} column densities for the ground level (open
  circle), first fine structure level (filled circle) and first excited
  level (upper limit) transitions for component I in the spectrum of
  GRB\,090926A. Column density predictions from our time-dependent
  photo-excitation code are also shown. They refer to the ground level
  (dotted line), first fine structure level (thick solid line) and first
  excited level (dashed line) transitions, in the case of an absorber
  placed at 2.6 kpc from the GRB. The two thin solid lines display the
  models which enclose the fine structure data at $1\sigma$ level. Bottom panel:
same as top panel, but for ground and first fine structure level transitions of \ion{Si}{II}.}
\label{spe1}
\end{figure}

To check the results from the steady state approximation, we must
compare our observed column densities to those predicted by a
photo-excitation code for the time when the spectroscopic observations
have been acquired. 
The photo-excitation code is that used by Vreeswijk et al. (2007) and
D'Elia et al. (2009), to which we refer the reader for further
details. Basically, it solves the detailed balance equation in a
time-dependent way for a set of transitions involving the levels of a
given species (e.g. \ion{Fe}{II} and \ion{Si}{II}).  The equation
depends on the flux level experienced by the absorbing gas. This flux
is of course a function of the distance $d$ of the gas from the GRB
explosion site, which is a free parameter of the computation and the
quantity we want to calculate. The other free parameters are the
initial column densities of the levels involved, which are assumed to
be in the ground state before the GRB flux reaches the gas, and the
Doppler parameter of the gas itself. The afterglow spectral index has
been calculated directly from the merged, flux-calibrated X-shooter
spectrum, and is $\beta \sim 0.9$ (see Sect. 6). The flux behaviour
before the X-shooter observation has been estimated using the data in
R10. In detail, if the flux in the $R$ band is $F_R=F_R(t_*) \times
(t/t_*)^{-\alpha}$, we have $F_R(t_*)=1.1\times 10^{-27}$ erg
cm$^{-2}$ s$^{-1}$ Hz$^{-1}$ at $t_*=72.9$~ks, with a decay index
$\alpha$ of 1.6 (-2.7) before (after) $t_*$. No spectral variation is
assumed during the time interval between the burst and our observation
(this assumption is strongly supported by the R10 data). The initial
column densities of the ground states have been computed from the
observed column densities of all the levels of each ion. The exact
values are: $\log (N_\ion{Si}{II}/{\rm cm}^{-2}) = 14.44 \pm 0.04$ and
$\log (N_\ion{Fe}{II}/{\rm cm}^{-2}) = 14.04 \pm 0.03$. Finally, the
Doppler parameter values used as input of this model are $30$ and $90$
km s$^{-1}$, i.e. the values that best fit the {\ion{Fe}{II}} and
{\ion{Si}{II}} absorptions corresponding to components I and II,
respectively.

We first modeled the {\ion{Fe}{II}} photo-excitation. Fig. 6 (top) shows
the model which best fits the {\ion{Fe}{II}} data and the two
theoretical curves compatible within the error bars for the first
{\ion{Fe}{II}} fine structure level column density. The distance of
component I from the GRB explosion site results $d=2.6\pm0.3$ kpc. The
same calculation has been performed using the {\ion{Si}{II}} atomic
data.  The results are displayed in Fig. 6 (bottom), and the estimated
distance is $d=2.25\pm0.15$ kpc, which is consistent with that
estimated using the {\ion{Fe}{II}} data. In addition, our distances
are less than 30\% away from that estimated using the code by Silva \&
Viegas (2002), and compatible with it at the $2\sigma$ level,
confirming that the steady state and optically thin approximations are
appropriate for component I of GRB\,090926A. Regarding component II,
we do not have {\ion{Fe}{II}} excited features, so we can just set a
lower limit to the GRB/absorber distance, which is $2.1$ ($2.2$) kpc
adopting a Doppler parameter of $30$ ($90$) km s$^{-1}$. The
{\ion{Si}{II}} fine structure features are instead present in
component II. The lack of {\ion{Fe}{II}} excited levels in components
featuring {\ion{Si}{II}} fine structure ones is not atypical, see
e.g. component III of GRB\,050730 identified in D'Elia et
al. (2007). Nevertheless, the large errors for this feature and the
strong dependence of the model from the Doppler parameter make the
distance estimation for this component quite uncertain. We run our
code using the Doppler parameters that best fit Components I ($30$ km
s$^{-1}$) and II ($90$ km s$^{-1}$).  The distance of component II
from the GRB is found to be $4.4\pm0.6$ kpc for $b=30$ km s$^{-1}$,
and $5.8\pm0.8$ kpc for $b=90$ km s$^{-1}$. Despite these values are
consistent with the lower limits estimated through {\ion{Fe}{II}}
data, we caution that a direct comparison to check the {\ion{Si}{II}}
distance is missing.

\section{Other features at the host redshift}

In this section we turn our attention to the search for galactic
emission lines (Sect. 5.1) and absorption features not due to matter in the
atomic gas phase. The latter can be produced by molecules (Sect. 5.2) or
DIBs (Sect. 5.3).

\subsection{Emission lines from the host galaxy}

We searched for most of the stronger emission lines like
[\ion{O}{ii}]~$\lambda$ 3727, [\ion{O}{iii}]~$\lambda$ 5007 and the
Balmer lines from the host galaxy in the X-shooter spectrum, but found
none. We did this by subtracting the spectral PSF. To derive upper
limits for e.g., H$\alpha$, we then added artificial emission lines of
increasing strength to the data, until the line was easily
detectable. For H$\alpha$, which in this case is observed at $2~\mu$m,
we find that an emission line of $9\times10^{-18}$
erg~s$^{-1}$~cm$^{-2}$ with an intrinsic velocity width of 100 km
s$^{-1}$ would have been detected (barring slit losses). This means
that we would have detected emission lines from a host galaxy with a
star formation rate $\gtrsim 2 M_\odot~\mathrm{yr}^{-1}$ (Kennicutt
1998). We also note that there is no Ly$\alpha$ emission in the trough
of the DLA. The 5$\sigma$ detection limit in a 6\AA{} wide extraction
window is $3.0 \times 10^{-17}$ erg~s$^{-1}$~cm$^{-2}$ corresponding
to $\mathrm{SFR} < 1 M_\odot$~yr$^{-1}$. However, this limit is much
more senstitive to uncertainties in radiative transfer of Ly$\alpha$
photons and to host galaxy dust extinction (see, e.g., Hayes et
al. 2010).  Although there is no evidence for dust extinction from the
SED of the afterglow radiative transfer can still completely remove
the Ly$\alpha$ emission from the line-of-sight (e.g., Atek et
al. 2008). We note that many GRB host galaxies do have higher SFR than
this limit (e.g., Christensen et al. 2004), and we therefore expect
to be able to detect emission lines from many high-z GRB hosts using
X-shooter in the future.



\subsection{Molecular absorption features}

Molecular gas is expected to be found in star-forming environments,
but the search for its absorption features has often given negative
results (see e.g. Tumlinson et al. 2007). Fynbo et al. (2006)
interpreted an absorption feature of the GRB\,060206 afterglow
spectrum as a possible H$_2$ detection. Prochaska et al. (2009) and
Sheffer et al. (2009) reported the presence of strong H$_2$ and CO
absorption features in the spectrum of the GRB\,080607
afterglow. These are the first (and by now unique) positive detections
of molecules in GRB host galaxies. For GRB\,090926A, we searched for
the strongest features of the CO molecule, e.g.,
\ion{CO}{}~$\lambda\lambda$1510, 1478 and 1447, but we found no
evidence of absorption at these wavelengths. Quantitatively, the upper
limit for the \ion{CO} column density depends on the adopted Doppler
value $b$. Assuming it is in the range $30 - 90$ km s$^{-1}$ (i.e.,
the values that best fits components I and II, respectively, see
Sect. 4.1), the resulting upper limit for the \ion{CO} column density
lies in the interval $\log (N_{\rm CO}/{\rm cm}^{-2}) < 14 - 14.3$.
We also searched for H$_2$ absorption. The strongest H$_2$ transitions
of the Lyman-Werner bands have a wavelength $<3000$~\AA~ at the
redshift of GRB\,090926A. Nevertheless, using the H$_2$ L2R0
transition, which falls around $3350$~\AA, we can set a $90\%$ upper
limit of $\log (N_{{\rm H}_2}/{\rm cm}^{-2}) < 14.9$ for a $b$
parameter fixed at $30$ km~s$^{-1}$ and $\log (N_{{\rm H}_2}/{\rm
  cm}^{-2}) < 15.3$ for $b$ fixed at $90$ km s$^{-1}$.  Finally, we
searched for other absorptions, due to the \ion{CH}, {\ion{CH$^+$}{}} and
{\ion{CN}{}} molecules, but also in these cases we found none.

\subsection{Diffuse interstellar bands}

DIBs (see Jenniskens \& Desert 1994 and references therein) are broad
absorption features in the near-infrared to UV wavelength range. They
were observed for the first time more than 70 years ago in the Milky
Way (Merril 1934), and more recently in near galaxies (see e.g., Cox
\& Cordiner 2008, Cordiner et al. 2008). Although they have been known
for a long time and hundreds of features have already been identified,
the nature of the carrier of such transitions is still uncertain. The
identification is difficult since most of them are not correlated with
each other. The most promising candidates are large molecules, which
are thought to be polycyclic aromatic hydrocarbons. We searched for
the ten strongest and most common DIBs in the GRB\,090926A afterglow
spectrum, in the rest-frame wavelength range $\lambda = 4000 -
7000$~\AA, but we found no positive detection. A conservative
$2\sigma$ upper limit to the rest frame equivalent width of these
systems (taking into account that they have very different FWHM) is
${\rm EW} < 0.3$~\AA. This limit is set by the signal-to-noise ratio
of the infrared arm of our X-shooter spectrum, i.e., is the ``noise
EW''.

\section{The extinction curve shape}

The shape of extinction curves at high redshift is a powerful tool to
derive information about dust formation and possibly about the various
processes affecting dust absorption and destruction close to GRB
sites. In most cases, inferences about dust extinction curves are
obtained by photometric observations of GRB afterglows which
unfortunately is biased by a strong degeneracy between afterglow
spectral slope and extinction. Moreover, in particular for very high
redshift events, uncertainties in photometry of single bands which can
also present specific calibration problems may affect the whole
analysis (see for instance discussion in Stratta et al. 2007 and Zafar
et al. 2010 about GRB\,050904). For early time afterglows even the
possible non-absolute simultaneity of the available photometric
information has to be properly considered to avoid spurious results.

For most GRB afterglows, when accurate multi-band photometry is
available (e.g., Covino et al. 2008, Schady et al. 2010, Kann et
al. 2010), the derived extinction curve is in fair agreement with what
observed locally in the Small Magellanic Cloud (SMC, Pei 1992)
although often, due to the limited wavelength resolution, this simply
means that the observed extinction curve has to be chromatic
(i.e. wavelength dependent) and featureless. A few remarkable
exceptions have been recorded in an intervening system along the line
of sight of GRB\,060418 (Vreeswijk et al. 2007), for GRB\,070802
(Kr\"uhler et al. 2008; El\'{\i}asd\'ottir et al. 2009) and
GRB\,080607 (Prochaska et al. 2009), where the characteristic
absorption feature at about 2175\,\AA, prominent in the Milky Way
extinction curve, has also been detected.

Clearly, when spectroscopic information is available
(e.g., Liang \& Li 2010), better constrained results can be derived,
disentangling the extinction curve and spectral slope effects. 

The case of GRB\,090926A is a good example of the X-shooter
capabilities in this context. The flux calibrated spectrum (Fig.\,1)
has been analyzed after removing wavelength intervals affected by
telluric lines of strong absorptions. The Ly$\alpha$ range has been
included in the analysis due to the importance of its wavelength range
for extinction determination using the Hydrogen column density
reported in Table\,2. We then rebin the spectrum in bins of
approximately 50\,\AA$\;$ by a sigma-clipping algorithm to avoid the
effect of residuals absorption systems. The flux-calibration of
X-shooter spectra ranging from the UV to the $K$ band is not an easy
task and the reduction pipeline (Goldoni et al. 2006) is still under
active development. In order to obtain an acceptable fit we had to
introduce additional systematic uncertainties at about $\sim 2.5$\%
level added in quadrature, which possibly reflects normalization
biases of the three arms and/or still not fully modeled slit losses
along the whole wavelength range.  Assuming a power-law spectrum, we
obtained $\beta = 0.89 \pm 0.02$ (errors at 1$\sigma$) where the
afterglow spectrum is modeled, as customary in GRB literature, as
$F_\nu \propto \nu^{-\beta}$. We try to fit the data using different
extinction curves, namely, SMC, LMC, Milky Way and Starburst. The best
fit is obtained assuming a SMC extinction curve with $E_{B-V} < 0.01$
mag at $3\sigma$ ($\chi^2/{\rm dof} = 1.14$ for 484 dof). Other
extinction curve recipes were not required by the data.

\section{The GRB\,090926A line of sight}

The absorption lines due to the gas belonging to the GRB host galaxy
are the dominant features of our spectrum, but they are not the only
ones. A detailed analysis of the data reveals that at least four other
absorbers are present along the line of sight to GRB\,090926A. Three of
these systems, those with the highest redshifts ($z=1.75 - 1.95$), show
absorption from the {\ion{C}{IV} $\lambda\lambda$1548, 1550} doublet,
to which corresponds a well defined {\ion{H}{I} $\lambda$1215} line
inside the Ly$\alpha$ forest. All these systems exhibit a very
simple line profile, that can be fitted by a single Voigt function
(Fig. 7). The Ly$\alpha$ of the system at $z=1.7986$ requires a
Doppler parameter that is about three times that for the \ion{C}{IV}
doublet, in order to be adequately fit. Anyway, if we assume a double
component model for this Lyman-$\alpha$ feature, and fix the $b$
parameter of one component to that of the \ion{C}{IV} doublet, the
estimated $N_{\rm H}$ column density is not significantly different from
that computed using the single component model. The last of the
GRB\,090926A intervening systems has a lower redshift ($z=1.2456$) and
features absorption from the {\ion{Mg}{II} $\lambda \lambda$2796,
2803} doublet and marginally ($2.7 \sigma$) from the {\ion{Mg}{I}
$\lambda$2852} line. This is a rather weak system, and the rest
frame equivalent width for the {\ion{Mg}{II} $\lambda$2796} line is
just EW$_{\rm r}=0.19\pm0.06~\AA$~($2\sigma$
confidence).  Again, a single component Voigt profile describes well
the absorption features (Fig. 7).
 
Table 4 summarizes the rest frame equivalent widths, column densities
and redshifts calculated for these intervening systems.

\begin{table}[ht]
\caption{\bf Redshifts, absorption line column densities and equivalent widths for the intervening systems.}
{\footnotesize
\smallskip
\begin{tabular}{|lcc|ccc|}
\hline
&Species        & Transition                    & Redshift    & $\log$ (N/cm$^{-2}$) & EW$_{\rm r}$ (\AA)$^a$  \\
\hline
1&CIV                 &  1548, 1550 & 1.9466  &  $   13.70 \pm 0.03 $  &  $ 0.15\pm 0.04 $  \\
\hline                                                                                               
1&HI                  &  Ly$\alpha$                  & 1.9466  & $   14.64 \pm 0.04$  &   $          $  \\
\hline                                                                                               
2&CIV                 &  1548, 1550 & 1.7986  &  $   13.63 \pm 0.03 $  &  $ 0.11\pm 0.03$  \\
\hline                                                                                          
2&HI                  &  Ly$\alpha$                  & 1.7986  & $   14.56 \pm 0.07 $ &   $          $  \\
\hline                                                                                          
3&CIV                 &  1548, 1550 & 1.7483  &  $   13.90 \pm 0.02 $  &  $ 0.21\pm 0.03$  \\
\hline                                                                                          
3&HI                  &  Ly$\alpha$                  & 1.7483  & $   14.98 \pm 0.41$  &   $          $  \\
\hline                                                                                          
4&MgII                &  2796, 2803 & 1.2456  &  $   12.39 \pm 0.05 $  &  $ 0.19\pm0.06$  \\
\hline                                                                                               
4&MgI                 &  2852                & 1.2456  &  $   11.47 \pm 0.13 $  &  $         $  \\
\hline
\end{tabular}

$^a$ Rest frame equivalent widths for {\ion{Mg}{II} $\lambda$2796} and {\ion{C}{IV} $\lambda$1548}; 
EW errors are given at the $2\sigma$ confidence level.
}
\end{table}

\begin{figure}
\centering
\includegraphics[angle=-0,width=4.4cm]{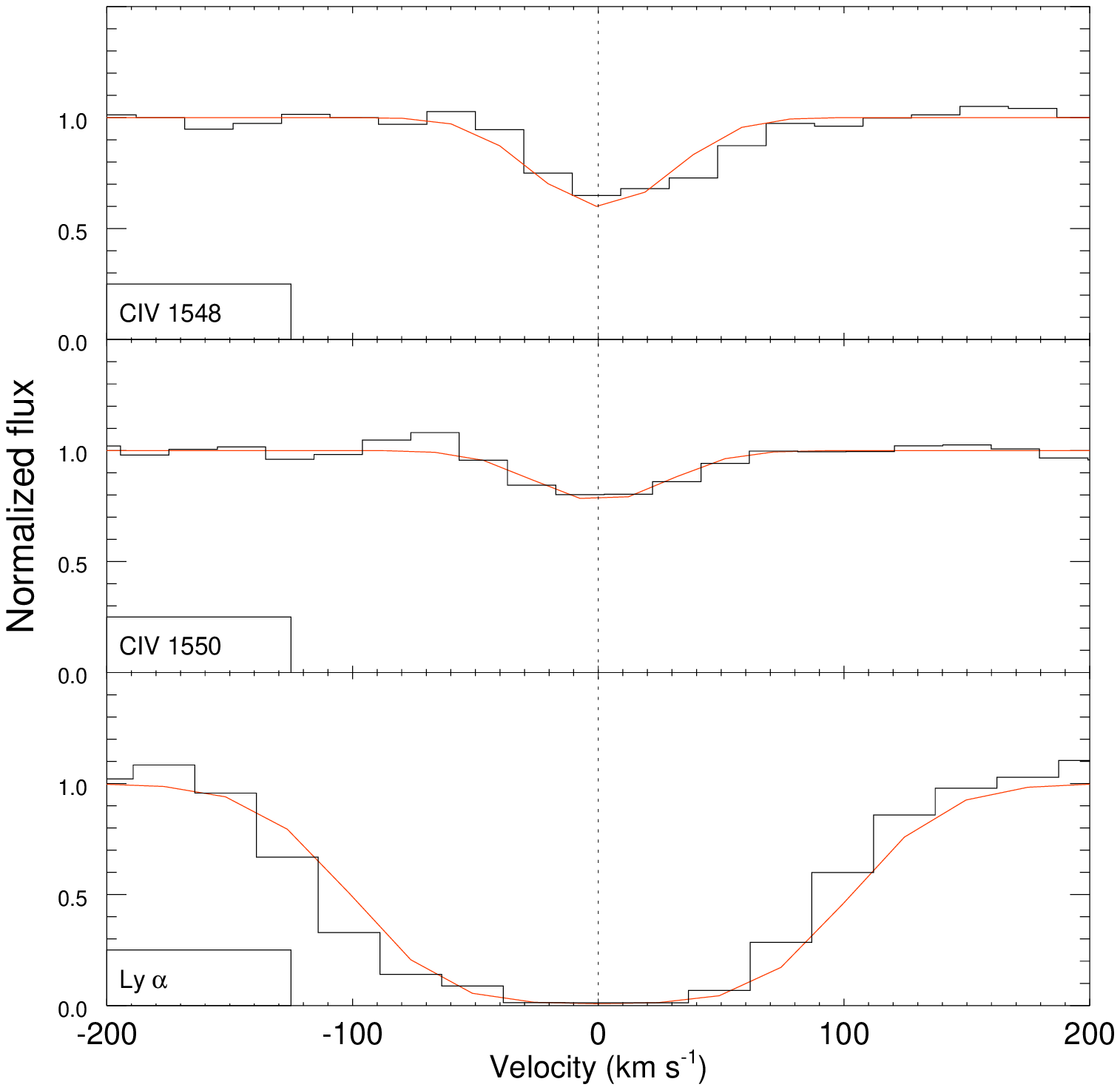}
\includegraphics[angle=-0,width=4.4cm]{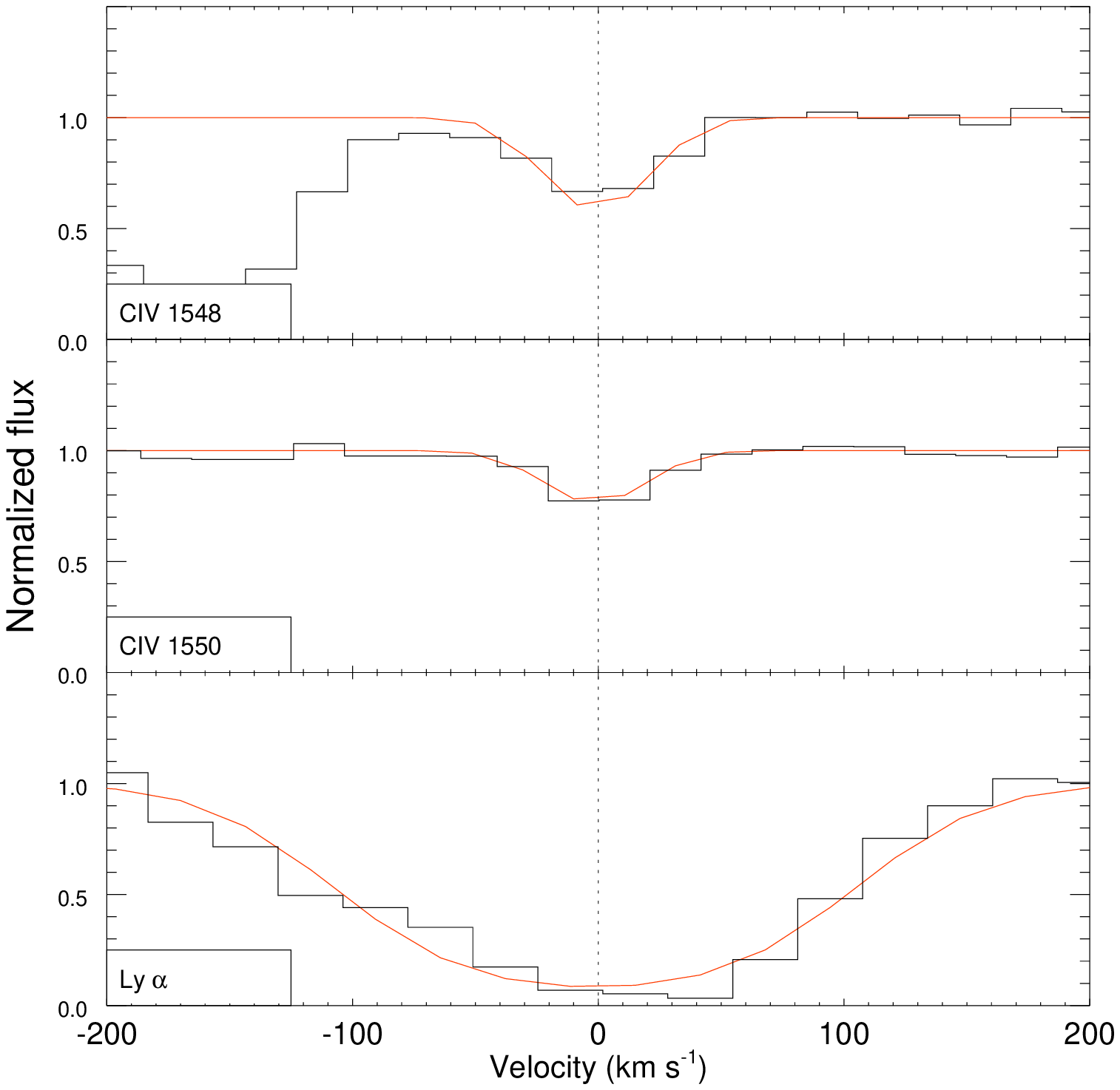}
\includegraphics[angle=-0,width=4.4cm]{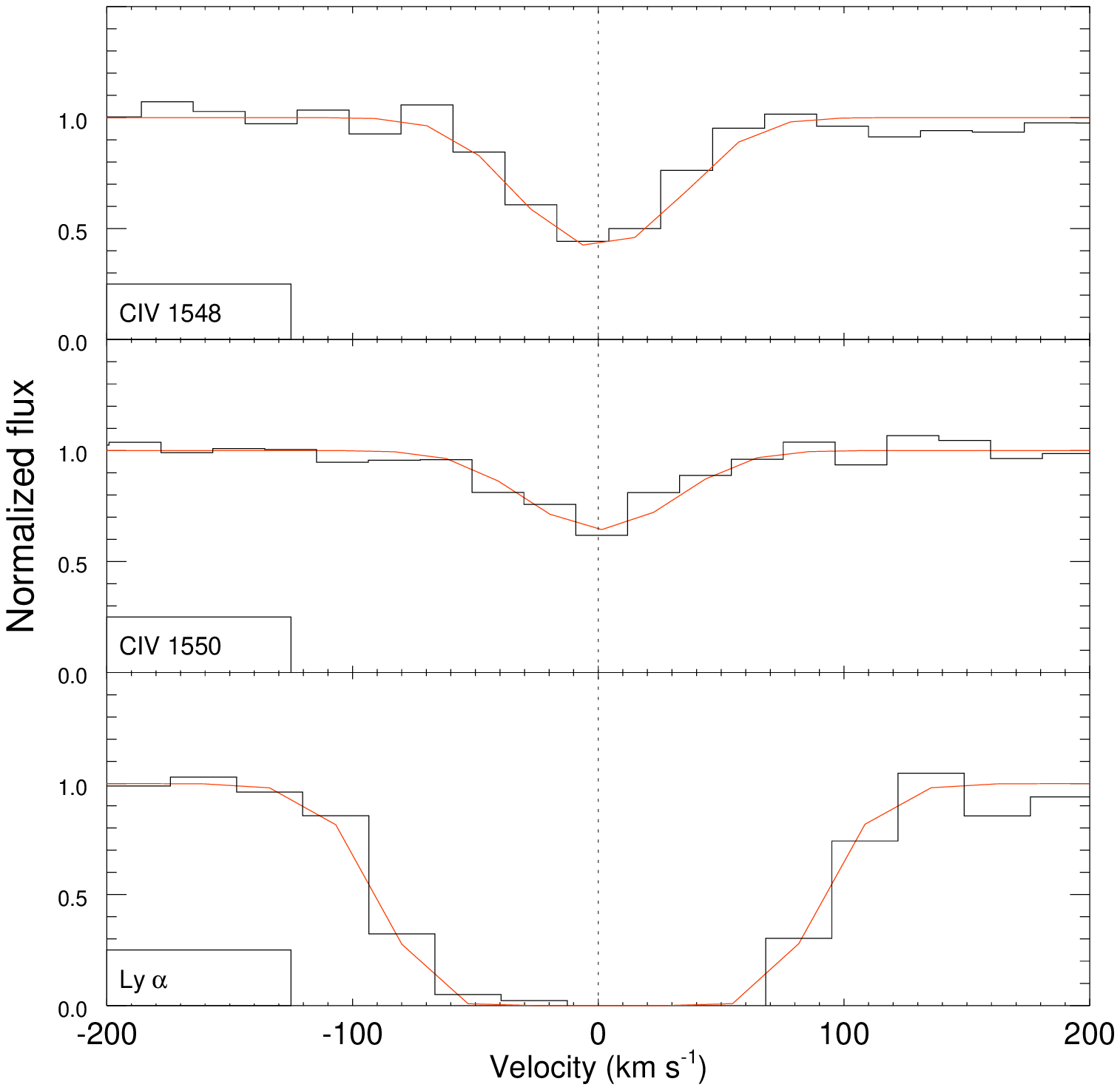}
\includegraphics[angle=-0,width=4.4cm]{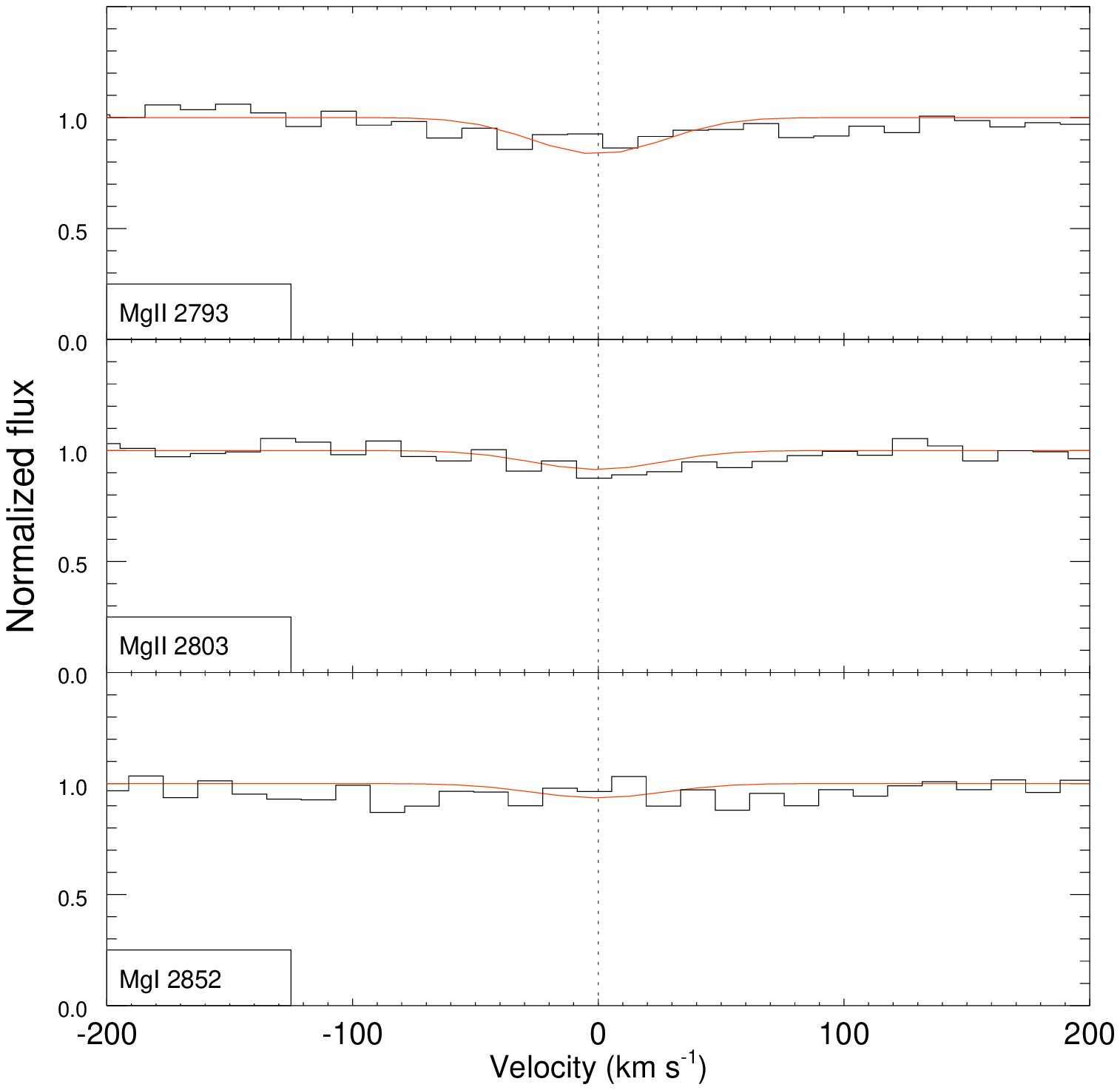}
\caption{The four intervening systems along the GRB090926A
  sightline. Top left: intervening 1 at $z=1.9466$; top right:
  intervening 2 at z=$1.7986$; bottom left: intervening 3 at
  $z=1.7483$; bottom right: intervening 4 at $z=1.2456$. The first
  three systems feature the \ion{C}{IV} doublet and Ly$\alpha$. System
  4 shows {\ion{Mg}{I}} and {\ion{Mg}{II}}, with the
  $\lambda$2852 transition and $\lambda$2796 - $\lambda$2803 doublet,
  respectively. {\ion{Mg}{I}} is marginally detected ($2.7
  \sigma$). Solid lines represent our best fit model (see also Table
  4).}
\label{spe1}
\end{figure}



                                                      

\section{Conclusions and discussion}

In this paper we present intermediate resolution ($R=10,000$)
spectroscopy of the optical afterglow of GRB\,090926A,
observed using the X-shooter spectrograph at the VLT $\sim 22$ hr
after the trigger.
                                                                                    
From the detection of Hydrogen and metal absorption features, we find
that the heliocentric redshift of the host galaxy is $z=2.1071$.  The
spectrum shows that the ISM of the GRB host galaxy has at least two
components contributing to this main absorption system at $z =
2.1071$. Such components, whose line centres are separated by $\sim
48$ km s$^{-1}$, are identified in this paper as I and II, according
to their decreasing velocity values. The total width of the two
components is $\sim 250$ km s$^{-1}$. We stress that the
identification of just two components may be due to the X-shooter
resolution. In fact, GRB afterglows observed with higher spectral
resolution feature more complex environments either if the host galaxy
absorber has a similar width (e.g., 7 components for GRB\,050922C, see
Piranomonte et al. 2008) or even in case of smaller widths (GRB\,080319B
has six components for a total width of $\sim 120$ km s$^{-1}$, see
D'Elia et al. 2009a).

The absorption lines appear both as neutral metal-absorption, low
ionization and high ionization species. In addition, strong absorption
from the fine structure and metastable levels of several species are
detected. The distances between the GRB and the two absorbers have
been estimated using the code by Silva \& Viegas (2002), in a steady
state and optically thin approximation, using the ratios between the
ground state and the first excited levels of different species to
infer the flux level experienced by the absorbing gas. For both
components we find that the absorber is located at a distance of
$d=1.8\pm0.2$ kpc.

The value of this distance can be refined by comparing the column
densities of the ground and excited levels to those predicted by a
time dependent photo-excitation code.  Using {\ion{Fe}{II}} and
{\ion{Si}{II}} we find that the absorbing gas of component I is
located at $d = 2.40 \pm 0.15$ kpc from the GRB, a value which is in
not far from that estimated assuming a steady state approximation. For
component II, this distance is larger, $\sim 5$ kpc, but this value
has been obtained using {\ion{Si}{II}} only, because we just have a
lower limit for the {\ion{Fe}{II}} fine structure column
density. Despite this lower limit gives a distance consistent to the
{\ion{Si}{II}} result, a safe cross check can not be performed.

The GRB\,090926A/absorber distance is compatible with that found for the
4 other GRBs for which a similar analysis has been performed, i.e.,
$1.7$ kpc for GRB\,060418 (Vreeswijk et al. 2007), $2-6$ kpc for
GRB\,080319B (D'Elia et al. 2009a), $280$ pc for GRB\,080330 (D'Elia et
al. 2009b) and $440$ pc for GRB\,050730 (Ledoux et al. 2009). This is a
further confirmation that the power of a GRB affects a region of gas
which is at least a few hundreds pc in size.

Several high ionization lines are detected in the GRB090926A spectrum
(see Table 2). Our {\ion{N}{V}} column density is within the range of
the GRB sample studied by Prochaska et al. (2008), who claim that this
ion can be located very close to the GRB explosion site. Fox et
al. (2008) estimated a lower limit to the {\ion{S}{IV}} distance from
the absorber using the non detection of the {\ion{S}{IV}} fine
structure feature and a photoexcitation code similar to that used in
this paper. They conclude that for GRB\,050730 this ion is located at
distances greater than $400$ pc. An upper limit of the same order of
magnitude can be roughly estimated also for GRB090926A, since we
compute a tighter limit for the {\ion{S}{IV*}} column density of $\log
(N_\ion{S}{IV}/{\rm cm}^{-2}) < 13.3$, but the flux of our GRB is
about three times smaller than that of GRB\,050730 (compared at the
time of the acquisition of the spectra). In addition, Fox et
al. (2008) report the detection in six of the seven GRB spectra
analyzed in their work of CIV high velocity components at
500-5000km/s. These gas can belong to foreground clouds but also be
associated to the Wolf-Rayet wind of the GRB progenitors. We do not
find any of such absorptions in the spectra of GRB\,090926A.

The redshift of GRB\,090926A allows us to determine the Hydrogen
column density, which has $\log (N_{\rm H}/{\rm cm}^{-2}) = 21.60 \pm
0.07$. This value is lower than that found by R10 ($\log (N_{\rm
  H}/{\rm cm}^{-2}) = 21.79 \pm 0.07$), but the $2\sigma$ regions
overlap, and each value is consistent with the other at the
$2.7\sigma$ level. Using $N_{\rm H}$ we evaluate the GRB\,090926A host
galaxies metallicity. The values we find are in the range
$10^{-3}-10^{-2}$ with respect to the solar abundances (see Table
3). R10 report a metallicity calculated using {\ion{S}{II}},
{\ion{Si}{II}}, {\ion{Al}{II}}, {\ion{O}{I}} and {\ion{Fe}{II}}. The
first three values are in perfect agreement with our estimates for S,
Si and Al. Their {\ion{O}{I}} value is consistent with our Oxygen
lower limit (set mainly by the saturated {\ion{O}{I}}
features). However, the authors note that their extremely low value
($\sim -3.08$) is possibly due to saturation and it could consequently
lead to an underestimate of the {\ion{O}{I}} column density, and thus
this should be regarded as a lower limit as well. Another way to
derive a metallicity using Oxygen, is to evaluate the {\ion{O}{I}}
ground state column density through the fine structure one, in the
hypothesis of a steady state UV pumping as the responsible for this
excitation. This has proven to be a good approximation for component
I, the only one featuring {\ion{O}{I}}. Under this assumption, we
obtain $\log (N_{\ion{O}{I}}/{\rm cm}^{-2}) \sim 16.05$ and a
metallicity of [O/H] $= -2.18\pm0.16$, which is consistent with the
values obtained using other elements. Finally, the R10 metallicity
obtained using the {\ion{Fe}{II}} features ([Fe/H] $\sim -2.93$) is
not consistent with our Iron value of [Fe/H] $= -2.19\pm0.07$. This is
because we considered the {\ion{Fe}{III}$\lambda$1122} line which
brings a relevant contribution to the total Iron column density for
GRB\,090926A. Due to this contribution, we do not detect an
overabundance of Silicon with respect to Iron, contrary to what
reported by R10.

For what concerns our results, the extremely low value derived for
\ion{N} ([N/H] $\sim -3.31$) is most certainly due to the fact that we
could not fit low ionization lines, but the \ion{N}{V} species
only. For \ion{Ca} ([Ca/H] $\sim -2.71$) we can not exclude a possible
contamination by sky lines. Despite this, we still have a metallicity
in the range $4.2\times10^{-3}$ - $1.4\times10^{-2}$ with respect to
solar. The average, logarithmic metallicity is [X/H] $= -2.14 \pm
0.09$ and is consistent with that computed by the R10 data excluding
Oxygen and Iron. This value lies at the lower end of the GRB
distribution (Savaglio 2006, Prochaska et al. 2007, Savaglio,
Glazebrook \& Le Borgne, 2009), in fact, only GRB\,050730 and
GRB\,050922C have metallicity values below $10^{-2}$.

A powerful way to infer the nature and the age of objects whose
morphology is unknown is to use abundances and abundance ratios (see
Matteucci 2001). This method is based on the fact that galaxies of
different morphological type are characterized by different star
formation histories and these strongly influence the [$X$/Fe] versus
[Fe/H] behaviour. For very large star formation rates, as expected in
spheroids (bulges and ellipticals) in their evolution phases, the
[$\alpha$/Fe] ratios are overabundant relative to the Sun in a large
range of [Fe/H] values, and this is because in a regime of strong star
formation the large number of Type II SNe acting in the early phases
of galaxy evolution increases the [Fe/H] in the gas up to large values
(almost solar and solar) on timescales too short for the Type Ia SNe
to substantially enrich the gas in Fe. Therefore, the [$\alpha$/Fe]
ratios will reflect the production ratio of SNe II which is larger
than in the Solar birthplace. In fact, the bulk of Fe is supposed to
have been produced by Type Ia SNe. On the other hand, objects evolving
slowly with a low star formation rate, with respect to normal
ellipticals, such as irregular galaxies, would show low [$\alpha$/Fe]
ratios even at low [Fe/H] values. This is because in this case SNe Ia
have time to pollute the gas much before it reaches the Fe solar
value.  Therefore, if we compare the measured abundance ratios in the
host of GRB\,090926A with predictions from detailed chemical evolution
models we can in principle understand the nature of the host. This is
done in Fig. 8 where models for ellipticals and for irregular
galaxies are compared with the measured abundances.  These models,
which are discussed in Fan et al. (2010, in preparation), show that
the host of GRB\,090926A is probably an irregular galaxy with
baryonic mass $10^{8}~M_{\odot}$ and evolving with a star formation
efficiency (the inverse of the timescale of star formation) of $0.05
{\rm Gyr}^{-1}$. In fact the measured [$\alpha$/Fe] ratios agree
better with the prediction for the irregular and are instead too low
for an elliptical galaxy of baryonic mass $10^{10}~M_{\odot}$, shown
for comparison. The model for the irregular galaxy takes into account
metal-enhanced galactic winds, induced by SN feedback, which produce
the loops in the predicted abundance ratios.

\begin{figure}
\centering
\includegraphics[angle=-0,width=9cm]{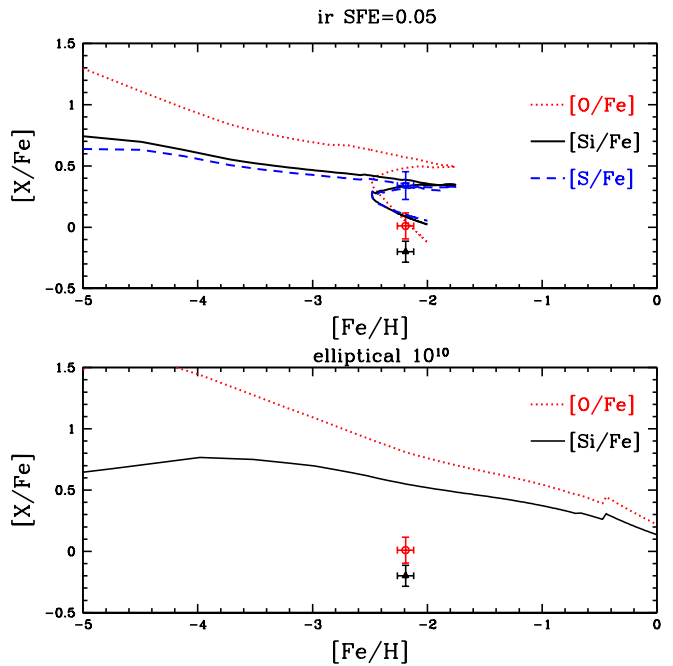}
\caption{Predicted [$X$/Fe] vs. [Fe/H] for a typical irregular galaxy
  (upper panel) with a baryonic mass of $10^{8}~M_{\odot}$ (star
  formation efficiency of 0.05 Gyr$^{-1}$) and for a small elliptical
  with baryonic mass $M=10^{10}~M_{\odot}$ (lower panel). The data for
  GRB\,090926A are shown for comparison.}
\label{spe1}
\end{figure}
 
This kind of host galaxy, associated with the low metallicity of the
GRB\,090926A ISM, agrees with the tentative evidence for a trend of
declining ISM metallicity with decreasing galaxy luminosity in the
star-forming galaxy population at $z = 2-4$ (Chen et al. 2009).

We also searched for other features at the host galaxy redshift.  No
emission lines were detected for the GRB\,090926A host, but our lower
limits ensure that we will be able to detect emission lines from many
high-redshift GRB hosts using X-shooter in the future.  Similarly, we
find no evidence for molecular absorption at the GRB redshift. The
lower limit on H$_2$ translates into an upper limit for the Hydrogen
molecular fraction of $f= 2N_{{\rm H}_2}/(2N_{{\rm H}_2}+N_{\rm H}) <
(3.0-7.4)\times10^{-7}$. The absence of molecules is not surprising,
since Hydrogen molecular fractions $\log f > -4.5$ are detected in
just $10\%$ of the QSO-DLA population. In addition, GRB environments
with metallicities below $0.1 Z_\odot$ (such as that of GRB\,090926A)
and low dust content can explain the lack of H$_2$ (see Ledoux et
al. 2009).  As a confirmation of this scenario, the only positive
detection of molecules in GRB host galaxies has been reported in the
GRB\,080607 afterglow, for which a solar or even super-solar
metallicity has been inferred (Prochaska et al. 2009).  Finally, DIBs
have not been detected either.

The GRB\,090926A continuum has been fitted assuming a power-law
spectrum. The best fit spectral index is $\beta =
0.89^{+0.02}_{-0.02}$ ($1\sigma$). This value is consistent with that
obtained in R10 by fitting the GROND data ($0.98^{+0.06}_{-0.07}$),
and is close to that obtained by the same authors fitting the
optical/IR and the X-ray data together ($\sim 1.03$). Our best fit
does not essentially allow for any intrinsic extinction since $E_{B-V}
< 0.01$ mag at $3\sigma$ adopting an SMC extinction curve. Other
extinction curve recipes are not required by the data. This intrinsic
extinction limit is consistent with that reported by R10.

The GRB\,090926A sightline has also been analyzed. The signal-to-noise
level of our spectrum allowed us to be sensitive to lines with
equivalent width (observed frame) as weak as $0.06-0.15$ \AA~
(depending on the spectral region), at the $2\sigma$ confidence
level. The redshift path analyzed for the search of \ion{Mg}{II}
(\ion{C}{IV}) systems ranges from $z=2.107$ to $z=0.35$
($z=1.44$). Four intervening systems between z = 1.95 and z = 1.24
have been identified. All systems have rest frame equivalent widths
very small, below $0.3$~\AA. Two of these systems, those marked as 1
and 3 in Table 4, have also been reported by R10, and their equivalent
widths are consistent with ours. The line of sight toward GRB\,090926A
is thus very clean, if compared to that of other GRBs.  In fact,
Prochter et al. (2006) and Vergani et al. (2009) claimed an excess of
strong (EW $>1$ \AA) {\ion{Mg}{II}} absorbers along GRB sight lines
with respect to QSO's. On the other hand, the number of \ion{C}{IV}
and weak \ion{Mg}{II} systems is consistent along the line of sight of
the two classes of objects (Tejos et al. 2007, 2009).  The reason for
the strong {\ion{Mg}{II}} discrepancy is still uncertain, despite
several possibilities have been already ruled out (e.g Porciani et
al. 2007, D'Elia et al. 2010). The only proposed explanation which is
still a possible candidate for this discrepancy is a multiband
magnification bias in GRB sightlines (Vergani et al. 2009). Anyway,
more observations and analysis are needed in order to solve this
issue.

\begin{acknowledgements}
  We thank an anonymous referee for several helpful comments in
  improving the quality and clarity of the paper. The Dark Cosmology
  Centre is funded by the Danish National Research Foundation.
 \end{acknowledgements}

\end{document}